%% file: paperLArSolarNu.tex
\documentclass[a4paper,11pt]{article}
\pdfoutput=1 
\usepackage{jcappub} 

\usepackage{graphicx}
\usepackage{color}
\usepackage{mathtools}
\usepackage[left]{lineno}
\usepackage{subfig}
\usepackage{hyperref}

\newcommand{\cpdt}{cpd/100 tonne}
\newcommand{\grs}{$\gamma$-rays}
\newcommand{\dsf}{DarkSide-50}

\usepackage{csquotes}
\usepackage{multirow}
  
\newcommand{\APC}{APC, Universit\'e Paris Diderot, CNRS/IN2P3, CEA/Irfu, Observatoire de Paris, Sorbonne Paris Cit\'e, Paris 75205, France}
\newcommand{\LPNHE}{LPNHE Paris, Universit\'e Pierre et Marie Curie, Universit\'e Paris Diderot, CNRS/IN2P3, Paris 75252, France}
\newcommand{\IPHC}{IPHC, Universit\'e de Strasbourg, CNRS/IN2P3, Strasbourg 67037, France}
\newcommand{\Davis}{Department of Physics, University of California, Davis, CA 95616, USA}
\newcommand{\UCLA}{Department of Physics and Astronomy, University of California, Los Angeles, CA 90095, USA}
\newcommand{\UMass}{Amherst Center for Fundamental Interactions and Department of Physics, University of Massachusetts, Amherst, MA 01003, USA}
\newcommand{\GEINFN}{Istituto Nazionale di Fisica Nucleare, Sezione di Genova, Genova 16146, Italy}
\newcommand{\GEUni}{Department of Physics, Universit\`a degli Studi, Genova 16146, Italy}
\newcommand{\NAINFN}{Istituto Nazionale di Fisica Nucleare, Sezione di Napoli, Napoli 80126, Italy}
\newcommand{\NAUni}{Department of Physics, Universit\`a degli Studi Federico II, Napoli 80126, Italy}
\newcommand{\Houston}{Department of Physics, University of Houston, Houston, TX 77204, USA}
\newcommand{ \AQGSSI}{Gran Sasso Science Institute, L'Aquila AQ 67100, Italy}
\newcommand{\AQLNGS}{Laboratori Nazionali del Gran Sasso, Assergi AQ 67010, Italy}
\newcommand{\Princeton}{Department of Physics, Princeton University, Princeton, NJ 08544, USA}
\newcommand{\MIINFN}{Istituto Nazionale di Fisica Nucleare, Sezione di Milano, Milano 20133, Italy}
\newcommand{\LSC}{Laboratorio Subterr\'aneo de Canfranc, Canfranc Estaci\'on 22880, Spain}
\newcommand{\Kurchatov}{National Research Centre Kurchatov Institute, Moscow 123182, Russia}
\newcommand{\Temple}{Department of Physics, Temple University, Philadelphia, PA 19122, USA}
\newcommand{\CAINFN}{Istituto Nazionale di Fisica Nucleare, Sezione di Cagliari, Cagliari 09042, Italy}

\newcommand{\pp}{\textit{pp}}
\newcommand{\pep}{\textit{pep}}
\newcommand{\hep}{\textit{hep}}
\newcommand{\lartpc}{LAr TPC}
\newcommand{\lartpcs}{LAr TPCs}

\title{Solar neutrino detection in a large volume double--phase liquid argon experiment}

\affiliation[a]{\APC}
\affiliation[b]{\LPNHE}
\affiliation[c]{\GEUni}
\affiliation[d]{\GEINFN}
\affiliation[e]{\Houston}
\affiliation[f]{\AQLNGS}
\affiliation[g]{\AQGSSI}
\affiliation[h]{\UCLA}
\affiliation[i]{\NAUni}
\affiliation[j]{\NAINFN}

\affiliation[k]{\Princeton}
\affiliation[l]{\MIINFN}
\affiliation[m]{\LSC}
\affiliation[n]{\IPHC}
\affiliation[o]{\Temple}
\affiliation[p]{\Davis}
\affiliation[q]{\UMass}
\affiliation[r]{\CAINFN}

\affiliation[s]{\Kurchatov}

\author[a,1]{D. Franco\note[1]{Corresponding author}}\emailAdd{dfranco@in2p3.fr}
\author[b]{C.~Giganti}\emailAdd{cgiganti@lpnhe.in2p3.fr}
\author[a]{P.~Agnes}\emailAdd{pagnes@in2p3.fr}
\author[b]{L.~Agostino}\emailAdd{lagostin@lpnhe.in2p3.fr}
\author[c,d]{B.~Bottino}\emailAdd{bianca.bottino@ge.infn.it}
\author[e,f]{N.~Canci}\emailAdd{nicola.canci@lngs.infn.it}
\author[f,g]{S.~Davini}\emailAdd{stefano.davini@gssi.infn.it}
\author[b]{S.~De Cecco}\emailAdd{sandro.dececco@lpnhe.in2p3.fr}
\author[h]{A.~Fan}\emailAdd{aldenf@physics.ucla.edu}
\author[i,j]{G.~Fiorillo}\emailAdd{giuliana.fiorillo@na.infn.it}
\author[k,l]{C.~Galbiati}\emailAdd{galbiati@princeton.edu}
\author[f]{A.~M.~Goretti}\emailAdd{augusto.goretti@lngs.infn.it}
\author[e]{E.~V.~Hungerford}\emailAdd{evhunger@central.uh.edu}
\author[f,m]{Al.~Ianni}\emailAdd{aldo.ianni@lngs.infn.it}
\author[k,f]{An.~Ianni}\emailAdd{ianni@princeton.edu}
\author[n]{C.~Jollet}\emailAdd{cecile.jollet@cern.ch}
\author[c,d]{L.~Marini}\emailAdd{laura.marini@ge.infn.it}
\author[o]{C.~J.~Martoff}\emailAdd{jeff.martoff@temple.edu}
\author[n]{A.~Meregaglia}\emailAdd{anselmo.meregaglia@cern.ch}
\author[c,d]{L.~Pagani}\emailAdd{luca.pagani@ge.infn.it}
\author[c,d]{M.~Pallavicini}\emailAdd{marco.pallavicini@ge.infn.it}
\author[p]{E.~Pantic}\emailAdd{pantic@ucdavis.edu}
\author[q,k]{A.~Pocar}\emailAdd{pocar@umass.edu}
\author[r]{M.~Razeti}\emailAdd{marco.razeti@ca.infn.it}
\author[h,e]{A.~L.~Renshaw}\emailAdd{arenshaw@central.uh.edu}
\author[j,k]{B.~Rossi}\emailAdd{biagio.rossi@na.infn.it}
\author[f]{N.~Rossi}\emailAdd{nicola.rossi@lngs.infn.it}
\author[f,h,s]{Y.~Suvorov}\emailAdd{yura.suvorov@lngs.infn.it}
\author[d]{G.~Testera}\emailAdd{gemma.testera@ge.infn.it}
\author[a]{A.~Tonazzo}\emailAdd{tonazzo@in2p3.fr}
\author[h]{H.~Wang}\emailAdd{hanguo@ucla.edu}
\author[d]{S.~Zavatarelli}\emailAdd{sandra.zavatarelli@ge.infn.it}

\date{\today}
\abstract{Precision  measurements of solar neutrinos emitted by specific nuclear reaction chains in the Sun are of great interest for developing an improved understanding of star formation and evolution. Given the expected neutrino fluxes and known detection reactions, such measurements require detectors capable of collecting neutrino-electron scattering data in exposures on the order of 1 ktonne-yr, with good energy resolution and extremely low background.  Two-phase liquid argon time projection chambers (\lartpcs) are under development for direct Dark Matter WIMP searches, which possess very large sensitive mass, high scintillation light yield, good energy resolution, and good spatial resolution in all three cartesian directions.  While enabling Dark Matter searches with sensitivity extending to the \enquote{neutrino floor} (given by the rate of nuclear recoil events from solar neutrino coherent scattering),  such detectors could also enable precision measurements of solar neutrino fluxes using the neutrino-electron elastic scattering events.

Modeling results are presented for the cosmogenic and radiogenic backgrounds affecting solar neutrino detection in a 300 tonne (100 tonne fiducial)  \lartpc\ operating at LNGS depth (3,800 meters of water equivalent). The results show that such a detector could measure the CNO neutrino rate with 
$\sim$15\% precision, and significantly improve the precision of the $^7$Be and \textit{pep} neutrino rates compared to the currently available results from the Borexino organic liquid scintillator detector.  

}

\begin{document} 

\maketitle
\flushbottom

\section{Introduction}
\input{Introduction}

\section{Neutrino signal and backgrounds}
\input{Neutrino}

\subsection{Cosmogenic background}
\input{Cosmogenics}

\subsection{Radon contamination}
\input{Radon}

\subsection{External background}
\input{ExternalBg}

\section{Sensitivity to solar neutrinos}
\input{Sensitivity}

\section{Conclusions}

\input{Discussion}

\acknowledgments
We acknowledge the financial support from the UnivEarthS Labex program of Sorbonne Paris Cit\'e (ANR-10-LABX-0023 and ANR-11-IDEX-0005-02). 


\input{Biblio}
\newpage


\appendix \section{Cosmogenic isotope production table}
\label{appendix}
\input{Appendix}

\end{document}

%% file: Introduction.tex
Fifty years of experimental study has yielded detailed information on neutrinos from the Sun, with profound implications for our understanding of nature. 
The Davis chlorine experiment \cite{Dav68} was the first to detect solar neutrinos, giving the historic proof that the Sun's energy is indeed produced by nuclear fusion reactions in its interior. Further study of the  \enquote{solar neutrino problem} (the overall deficit of solar neutrinos detected on Earth by Davis and subsequent experiments [GALLEX/GNO \cite{Ham98,Alt00}, SAGE \cite{Abd09}, Kamiokande \cite{Fuk96}, and SuperKamiokande \cite{Fuk98}]) led to the paradigm-changing 2001 discovery of ``neutrino flavor oscillations," which imply nonzero neutrino masses.  This occurred as the Sudbury Neutrino Observatory (SNO) experiment \cite{Aha05} showed that the solar neutrino problem was explained by  electron-type neutrinos from the Sun changing in transit into the other two active neutrino flavors (muon- and tau-type), which cannot undergo the low energy charged-current detection reactions used by all preceding experiments. Soon after,  the LMA-MSW \cite{MSW} oscillation paradigm for the neutrino oscillation phenomenon was confirmed by the KamLAND experiment using reactor electron-antineutrinos \cite{Egu03}. In 2007, the Borexino experiment \cite{Arp08b,Arp08,Bel10b,Bel11,Bel12} began a new phase of solar neutrino measurements intended to make precision tests of the  Standard Solar Model (SSM)~\cite{Bel14} and important astrophysics \cite{Bel15} and geophysics \cite{Bel10} measurements. 

To date, spectral features corresponding to almost all solar neutrino production reactions have been observed, with the notable exception of neutrinos produced in the $^{13}$N, $^{15}$O, and  $^{17}$F reactions  of the CNO cycle.   The CNO cycle   plays a key role in astrophysics, since it is  the dominant source of energy in stars more massive than the Sun and in advanced evolutionary stages of Sun-like stars. Solar CNO neutrino measurements would constrain the chemical composition of the Sun, leading to improved models  for star formation and supernova explosions.  A direct measurement of the CNO neutrino components could also solve the long-standing and well--known \enquote{solar metallicity problem}.   
The solar metallicity (fraction of elements with Z$>$2) is an important parameter in the SSM.  The metallicity is inferred from spectroscopy, and when input to SSM calculations its value affects the predicted sound speed radial profiles within the Sun (accurately measured  via helioseismology), and also changes some neutrino fluxes \cite{SBF09,SHP11}.  Two ranges of metallicity, known as the low-metallicity \cite{AGS09} and high-metallicity solutions \cite{GS98}, have been derived from spectroscopic measurements of the Sun.  The newest and recommended range (that of Ref.~\cite{AGS09})  results in a stark disagreement between predicted and observed sound speed profiles \cite{SHP11}.  
As shown in Table \ref{tab:fluxes}, the CNO neutrino flux predictions discriminate between the two abundance ranges.  A direct measurement of the CNO neutrino flux with a 10--20\% uncertainty could solve the solar metallicity problem.

\begin{table}
\begin{center}
\begin{tabular}{lccccccccc}
\hline
Neutrino &  & \multicolumn{3}{c}{GS98}  &  \multicolumn{3}{c}{AGSS09} & $\Delta$ \\
 Source &  Unit & Flux & +$\sigma$ & -$\sigma$  	&	Flux & +$\sigma$ & -$\sigma$ &  \\
\hline				                                     
\pp    & $\times$10$^{10}$&5.97   &+0.006 &-0.006     &  6.03  & +0.005 & -0.005   & -1.0\%     \\
\pep   &$\times$10$^{8}$&1.41   &+0.011 &-0.011     &  1.44  & +0.010 &-0.010       & -2.1\% \\
\hep   &$\times$10$^{3}$&7.91   &+0.15  &-0.15       & 8.18  & +0.15  &-0.15         & -3.3\% \\
$^7$Be&  $\times$10$^{9}$&5.08   &+0.06  &-0.06       & 4.64  & +0.06  &-0.06      & 9.1\%   \\
$^8$B    &$\times$10$^{6}$&5.88   &+0.11  &-0.11       & 4.85  & +0.12  &-0.12      & 19.2\%    \\
$^{13}$N  &$\times$10$^{8}$&2.82   &+0.14  &-0.14      &  2.07  & +0.14 & -0.13   & 30.2\%    \\
$^{15}$O  &$\times$10$^{8}$&2.09   &+0.16  &-0.15       & 1.47   &+0.16  &-0.15    & 34.8\%     \\ 
$^{17}$F   &$\times$10$^{6}$&5.65   &+0.17  &-0.16       & 3.48   &+0.17  &-0.16   & 47.5\%      \\
\hline
\end{tabular} 
\end{center}
\caption{Solar neutrino fluxes predicted by the GS98 high--metallicity \cite{GS98} and  AGSS09 low--metallicity \cite{AGS09} solutions with the Standard Solar  Model. Units are in cm$^{-2}$ s$^{-1}$. The discrepancy $\Delta$ is evaluated as: $2 \times (GS98-AGSS09)/(GS98+AGSS09)$. }
\label{tab:fluxes}
\end{table}

The most stringent experimental limit on the CNO  neutrino rate is from Borexino  \cite{Bel14} using elastic neutrino-electron scattering.  They obtained an upper limit for the interaction rate of  7.9 counts per day per 100 tonnes (\cpdt) at the 95\% C.L., which corresponds to a flux limit of $<$7.9$\times$10$^6$~cm$^{-2}$~s$^{-1}$, nearly double the expectations from Table \ref{tab:fluxes}. Borexino is a large organic liquid scintillation detector with excellent intrinsic radio-purity.  Although the $^{210}$Bi activity in the active mass of Borexino  is exceptionally low ($\sim$20 \cpdt), its spectral shape is very  similar to the expected CNO signal.  This precludes a positive neutrino signal from being extracted. For this reason, the Borexino collaboration is planning a new purification campaign, mainly focused on $^{210}$Bi removal. 

Liquid argon (LAr)   presents an excellent alternative  for observing CNO neutrinos via neutrino-electron elastic scattering.  LAr is a powerful scintillator, 4-5 times brighter than organic liquid scintillators.  In addition LAr, as a liquefied noble element, does not react and does not bond with chemical species, and can be maintained with very high purity in either the liquid or the gas phase.  The ultimate impurity levels of  $^{238}$U and $^{232}$Th achievable in LAr are lower than those achieved in organic liquid scintillators \cite{Bel14}.  The main impurity background is dissolved radon, emanating from vessels, pipes, and filters. But thanks to the low normal boiling temperature of LAr, it has been demonstrated that radon can be removed to a level of  $\sim$$\mu$Bq/m$^3$ gas by cryogenic adsorption on activated carbon \cite{Sim09}.

Two-phase \lartpcs\ have an LAr target volume and a thin gas layer at the top, where the ionization electrons surviving recombination are extracted and accelerated to produce a secondary light signal by gas proportional scintillation (GPS).  \lartpcs\ can determine the position of energy deposit events in the liquid to within a few mm in the drift direction, and $\sim$1 cm or better in the two transverse directions.  For a program of precision measurements on solar neutrinos, use of an \lartpc\ with fiducial mass of several hundred tonnes would bring a number of important advantages.  The fiducial volume could be sharply defined, removing backgrounds from surfaces and eliminating the largest source of systematic error currently affecting $^7$Be measurements. It would also allow strong suppression of  gamma-ray background by identifying and rejecting multiple-Compton scatters and other events having multiple energy deposition sites.  

DarkSide-50 \cite{Agn15} is a 50 kg two--phase \lartpc\ designed to search for nuclear-recoil events from galactic WIMP scattering. Recent \dsf\ results demonstrate the excellent performance of this technology for discriminating nuclear-recoil from electron-recoil events, and in radio-purity of the detector and active medium. On the basis of these results, the DarkSide collaboration is proposing a roadmap for the construction of  an LAr WIMP Dark Matter search detector able to reach the so--called \enquote{neutrino floor}.  This is the  Dark Matter sensitivity level set by  the irreducible nuclear-recoil background from coherent nuclear scattering of solar and atmospheric neutrinos   on argon nuclei \cite{Gal15}.  The projected exposure of 1,000 tonne yr would allow more than 10,000 CNO neutrino-electron scattering events to be collected.  Such a detector would also have  the capability of observing other components of the solar neutrino spectrum, such as $^7$Be and \textit{pep}, with a more favorable  signal--to--background ratio than  Borexino. These neutrino-electron scattering events would not be a background problem for the WIMP search, due to the excellent particle identification capability of the \lartpc.

The present work   explores the  potential for solar neutrino measurements from  the large mass LAr detectors being  designed for direct Dark Matter searches.  In particular, we investigate the expected measurement precision for neutrino rates as a function of the background contaminations, taking into account the detector response and associated systematics.

%% file: Neutrino.tex
The overall  solar neutrino-electron scattering  rate expected in LAr is $\sim$150 \cpdt. However, the spetral range below $~$0.6 MeV is inaccessible, due to background from $^{39}$Ar, (Q($\beta  ^- $)= 0.565 MeV, t$_{1/2}$ =269 y) produced by cosmic ray spallation of $^{\rm{nat}}$Ar. Atmospheric argon with its $\sim$9 $\times$10$^9$ \cpdt\ ($\sim$1~Bq/kg) of $^{39}$Ar is prohibitively radioactive for use in any large \lartpc. Argon extracted from underground gas wells (UAr) has been shown by the Darkside collaboration to contain only  $\sim$6 $\times$10$^6$ \cpdt\ (0.7 mBq/kg)  of $^{39}$Ar~\cite{Agn15b}.  But even this level would prevent extraction of solar neutrino signals in the low energy region, which is dominated by \textit{pp} neutrinos. However, $^7$Be neutrino interactions, whose Compton-like edge is expected at $\sim$0.66 MeV, would be accessible (particularly with UAr) thanks to the excellent energy resolution achievable in LAr.

The 3-D event localization available in an \lartpc\ allows the pulse height of single-site events to be fully position corrected, which should result in an energy resolution limited by photoelectron statistics.  With a  scintillation photon yield of  $\sim$40,000 photons/MeV \cite{Dok90}, LAr is approximately a factor 4 brighter than  organic liquid scintillators.   DarkSide-50 \cite{Agn15} has already demonstrated   a photoelectron yield of $\sim$7000 photoelectrons (PE)/MeV at a 200 V/cm field ($\sim$8500 pe/MeV at zero field), albeit in a relatively small detector.    MicroCLEAN \cite{Lip10}, a small, single-phase LAr detector, measured a  yield of  $\sim$6000 PE/MeV at zero field, constant within 2\%  in the [0.04, 0.66] MeV energy range.  Such  yields are approximately 12-14 times that achieved in Borexino ($\sim$500 PE/MeV), the solar neutrino experiment with the best resolution ever reached. 

Light yield, however, can be strongly  affected when scaling  LAr target masses from a few tens of kilograms (DarkSide-50 and MicroCLEAN) to hundreds of tons. Optical effects in the UV range, like  Rayleigh scattering and  light absorption, can, in fact, reduce the light yield when photon travels  long distances. Different techniques were used to estimate the Rayleigh scattering length: by calculation (90~cm \cite{Sei02}), direct measurement (66 $\pm$ 3~cm \cite{Ish97}), and by extrapolation from measurements at higher wavelengths (55 $\pm$ 5~cm \cite{Gra16}). The absorption length is expected to be much larger, but, since the light propagation is dominated by the Rayleigh effect, a direct measurement is extremely challenging. Theoretically, it has been estimated  \cite{Wri11} that, as a consequence of the Stokes' shift,  a LAr dimer should be in a vibrational excited state almost 1~eV above the ground state   to re-absorb the emitted photon. Such a state can be formed only during atom collisions, making the LAr self--absorption negligible. The dominant process in the light absorption in LAr is expected then to be strongly dependent on the presence of chemical impurities or noble gases like nitrogen, which can  affect the attenuation length  already at the ppb level \cite{Jon13}.  Such impurities can be efficiently removed via adsorption or via distillation. The DarkSide collaboration is building the 350 m tall ARIA   \cite{Gor15} cryogenic distillation column to assess the feasibility of isotopic separation of $^{39}$Ar from $^{40}$Ar.  This facility will have unprecedented capabilities for removal of chemical impurities and could have a significant impact in improving the optical properties of LAr.

\begin{figure}
\begin{center}
\includegraphics[width=1\linewidth]{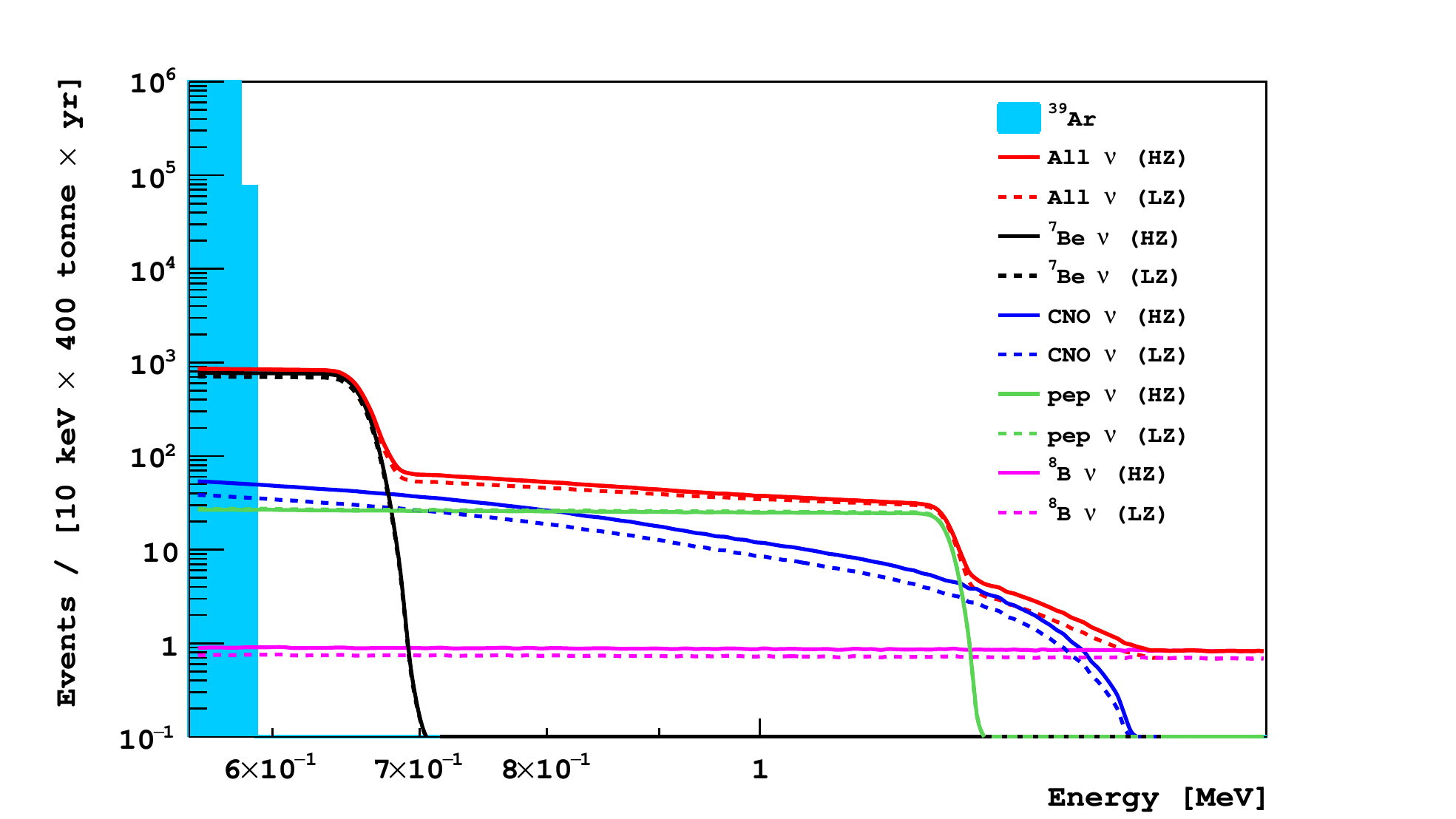}
\caption{Simulated solar neutrino spectra in a 400 tonne-yr \lartpc\ exposure, assuming $\sigma=$1.3\% energy resolution at 1 MeV, corresponding to a PE yield of  6 PE/keV.  The blue shaded area represents the tail of the  $^{39}$Ar contamination, intrinsic to underground LAr.}
\label{fig:nuspectra}
\end{center}
\end{figure}

Assuming photoelectron-statistics-limited resolution with 6000 PE/keV, and an $^{39}$Ar specific activity of 0.7 mBq/kg, no event from $^{39}$Ar is expected to appear above 0.6 MeV due to resolution smearing in a 400 tonne-year exposure (see Figure \ref{fig:nuspectra}). The present  paper will therefore consider the energy range  of interest for solar neutrinos to have a threshold at 0.6 MeV. 

Other possible background sources are $^{42}$Ar (Q($\beta  ^- $)= 0.599 MeV, $\tau_{1/2}$=33 yr), and its daughter $^{42}$K (Q($\beta  ^- $)= 3.52 MeV, $\tau_{1/2}$=12.4 hr).  Recent measurements by the GERDA collaboration \cite{Cat12} with atmospheric argon gave a  $^{42}$Ar specific activity of $\sim$8 $\times$10$^5$ \cpdt\ (94.5$\pm$18.1 $\mu$Bq/kg), which would presumably be in equilibrium with the  $^{42}$K daughter at the same level.   $^{42}$Ar can be  produced by two sequential neutron captures on $^{40}$Ar (mostly  during atmospheric nuclear tests)
or from  spallation by cosmic ray $\alpha$-particles ($^{40}$Ar($\alpha$, 2p)$^{42}$Ar) \cite{Bar97}. (Cosmic $\alpha$'s amount to  14\% of the cosmic proton flux.) Underground argon is not  subject to either of these mechanisms, and so would be expected to contain a much lower level of $^{42}$Ar than atmospheric argon.  

Table \ref{tab:nuspectra} shows the expected solar neutrino rates in LAr from $^7$Be, $^8$B, \textit{pep}, and CNO neutrinos in the  [0.6, 1.3]~MeV energy range, comparing the low--metallicity (LZ) solution~ \cite{AGS09} with the high--metallicity  (HZ) solution~\cite{GS98}. Rates for both solutions are calculated with  the  SSM and the neutrino oscillation survival probabilities from the MSW-LMA (large mixing angle) solution, using $\Delta m^2$ = 7.54$\times$10$^{-5}$~eV$^2$ and sin$^2$\   ($\theta_{12}$) = 0.307~\cite{MSW}. For the LZ solution the total neutrino rate is 4.63 $\pm$ 0.22 \cpdt, which increases to 5.14$\pm$0.25 \cpdt\ for the HZ solution. The main contribution  ($\sim$3 \cpdt) is from $^7$Be, even though this contributes only in  the range [0.6, 0.7]~MeV, as shown in Figure \ref{fig:nuspectra}. The HZ solution predicts  almost equal contributions from  \textit{pep} and CNO neutrinos above 0.7 MeV, while in the LZ model the \textit{pep} rate in this range is expected to be twice as large as that from CNO neutrinos. The $^8$B component is not  measurable  in the [0.6, 1.3]~MeV energy range, since its contribution is very small and featureless. $^8$B may be observable at high energies ($>$5~MeV) where  radioactive backgrounds and other contributions are negligible, but this is beyond the scope of this work and will be not treated.  

Aside from the argon isotopic impurities discussed above, the main backgrounds in the  [0.6, 1.3]~MeV energy range for a detector operating at LNGS depth are expected to be: 
\begin{itemize}
\item cosmogenic radionuclides produced in the running detector, by the interaction of cosmic rays with LAr;
\item   $^{222}$Rn dissolved in the LAr after being emanated from the detector walls or from the recirculation system;
\item external background, \textit{i.e.} \grs\ from radioactive contaminants in the detector construction materials.
\end{itemize}

\begin{table}
\begin{center}
\begin{tabular}{lcccc}
\hline
\multirow{2}{*}{Neutrino Source}          &   \multicolumn{2}{c}{Low Metallicity (LZ)} &    \multicolumn{2}{c}{High Metallicity (HZ)}   \\ 
     &  All &   [0.6-1.3]~MeV  & All &   [0.6-1.3]~MeV    \\ 
\hline
\textit{pp} &  107.9 $\pm$ 2.0 & 0 &107.0 $\pm$ 2.0& 0\\
\textit{pep} &2.28 $\pm$ 0.05&  1.10 $\pm$ 0.02 & 2.23  $\pm$ 0.05   &1.07 $\pm$ 0.02 \\
$^7$Be &   36.10 $\pm$ 2.60 & 2.85 $\pm$ 0.21  &39.58 $\pm$ 2.85 & 3.13 $\pm$ 0.23 \\
CNO &  3.06 $\pm$ 0.30 & 0.64 $\pm$  0.06 &4.28 $\pm$ 0.44& 0.90 $\pm$ 0.09\\
$^8$B& 0.30 $\pm$ 0.04  &  0.035  $\pm$ 0.005 & 0.36 $\pm$ 0.06 & 0.042  $\pm$ 0.007\\
\hline
Total& &  4.63 $\pm$ 0.22 &  &5.14 $\pm$ 0.25\\
\hline
\end{tabular} 
\caption{Expected solar neutrino rates in \cpdt\ of LAr active mass, comparing the low-metallicity \cite{AGS09} and high-metallicity \cite{GS98} predictions using the Standard Solar Model and neutrino oscillation parameters from the  MSW-LMA  \cite{MSW}  region with $\Delta m^2$ = 7.54$\times$10$^{-5}$~eV$^2$ and sin$^2$($\theta_{12}$) = 0.307.}
\label{tab:nuspectra}
\end{center}
\end{table}

Most of these backgrounds produce \grs, which allows them to be strongly suppressed by the multi-site interaction discrimination available in an \lartpc. If an event produces two or more interactions (multiple Compton scatters, etc.) separated by just a few millimeters along the drift direction, the ionization electrons from each interaction reach the gas layer several microseconds apart, generating  secondary gas proportional scintillation pulses well separated in time. For instance, assuming an electric field of 200 V/cm, the  drift velocity in LAr is $\sim$1~mm/$\mu$s. Two interactions  occurring 1 cm apart  along the drift direction will generate GPS light pulses separated by about 10~$\mu$s.  The characteristic scintillation emission times of gaseous argon ($\tau_{fast}$ $\sim$ 6~ns and $\tau_{slow}$ $\sim$ 1.6~$\mu$s)  allow the start times of pulses to be measured with a precision of tens of nanoseconds, thus easily allowing identification of events with even rather closely-spaced multiple interactions. The resulting discrimination factors are estimated by Monte Carlo codes developed and benchmarked in connection with the \dsf\ experiment.

Another possible source of background would be  $^{85}$Kr (Q($\beta  ^- $)= 0.687 MeV, $\tau_{1/2}$=10.75 yr) in the LAr fill.
The  DarkSide collaboration has measured a $^{85}$Kr  contamination in underground argon  of $\sim$1.7$\times$10$^7$ \cpdt\ (2 mBq/kg \cite{Agn15b}).  
$^{85}$Kr  in UAr can originate from atmospheric leaks or deep underground natural fission processes \cite{Agn15b}, which are known to give $^{85}$Kr activity in deep underground water reservoirs \cite{Leh10}.  \\

Despite the fact that only $\sim$1\%  of $^{85}$Kr decay events have an energy above 0.6 MeV, a contamination of the order of that measured in UAr by \dsf\ would seriously compromise the $^7$Be neutrino rate measurement.  However, it would not affect the measurement of higher energy CNO neutrinos. The DarkSide collaboration is planning to fill the planned multi-ton detector using argon purified by cryogenic distillation in columns with thousands of equilibrium stages~\cite{Aria}, a technique that can reduce the $^{85}$Kr activity well below  10 \cpdt\ (100 $\mu$Bq/(100 tonne) \cite{Dav15}).  This would effectively eliminate the $^{85}$K interference with the $^7$Be measurement.

In the next subsections, the contamination from each background source, and the associated residual rate after the multiple scattering cut,  will be discussed in detail.

%% file: Cosmogenics.tex
Direct dark matter search experiments are  located  deep underground in order to be shielded against the cosmic rays and their interaction products. At moderate depths however, the residual muon flux may still produce measurable amounts of radioactive isotopes by muon--induced spallation on the argon.  These can produce dangerous, delayed electron-recoil  background in the solar neutrino energy window.

Radiochemical and scintillator experiments for solar neutrino physics, like GALLEX and Borexino, have investigated   muon-induced radionuclide production using dedicated setups exposed to muon beams at CERN \cite{Cri97, Hag00}.  Further, radionuclide production models in simulation packages like GEANT4 \cite{Geant4} and FLUKA \cite{Fluka} have been greatly improved in recent years.  Predictions now typically agree within a factor of $\sim$2 with measured values, as shown by the Borexino  in \cite{Bel13} and  KamLAND collaborations  \cite{Abe10}. The present work uses FLUKA for evaluating the  cosmogenic production rates, since it has been found to be in better  agreement with  the Borexino measurements \cite{Bel13}.

 The detector geometry simulated in the present FLUKA code was a cylindrical TPC of 3.3~m radius and 3~m height, corresponding to a $\sim$150 tonne LAr active mass. The TPC sidewall is a 3~cm thick teflon layer, and the top and bottom ends are covered with 2~mm thick silicon layers, representing  silicon photomultiplier arrays. The 2 cm thick gaseous argon region sits just below the upper silicon layer. The TPC is contained in a 3~mm thick cylindrical stainless steel  cryostat with 3.5~m radius and 3.2~m height. Gaseous argon is also present in the cryostat, outside  the TPC region and above the LAr level.  The cryostat is  housed in a Borexino-like veto detector consisting of a 6~m radius stainless steel sphere filled with liquid scintillator placed within a larger cylindrical tank (17~m height, 16~m diamater) filled with water. In this work, the veto is considered as a passive shield only, despite the fact that the detection of light in the scintillator can provide a powerful rejection tool against external backgrounds.  As in the cosmogenic  simulation study for Borexino described in \cite{Bel13}, we used as input the cosmic ray muon flux measured by  MACRO at Gran Sasso laboratory ($\langle E_{\mu} \rangle$ = 283~GeV,  1.14 $\mu$/(hr m$^2$) \cite{Amb95}), and a $\frac{\mu^+}{\mu^-}$ ratio of 1.38, as measured by the OPERA experiment \cite{Aga10}.

 Cosmic muons were generated 3~m above the cylindrical water tank with an intervening   0.7~m  rock layer, in order to  fully develop the hadronic showers. The radionuclide production rates were then converted to specific activities, taking into account the mean life of each isotope.
 The simulation resulted in the production of  more than 80 isotopes  by muon spallation on argon,  as shown   in the tables in Appendix \ref{appendix}.  Cosmogenic isotopes produced in the other materials of the detector are taken into account in this analysis, but do not significantly  contribute  to the overall background.

\begin{table}
\begin{center}
\begin{tabular}{lccccc}
\hline
\multirow{2}{*}{Isotope}  & \multirow{2}{*}{Half Life} &   \multirow{2}{*}{Decay Mode} & Q-value  & \multicolumn{2}{c}{Rate}\\    
  & & & [MeV] &   Entire  Range &   [0.6-1.3]~MeV     \\ 
\hline
$^{41}$Ar &  109.61 min & $\beta^-$ &    2.492     & 0.213  & 0.054 \\
$^{38}$Cl &  37.230 min & $\beta^-$ &    4.917     & 0.815  & 0.147 \\
$^{39}$Cl & 55.6 min & $\beta^-$ &    3.442   &   0.173 & 0.051 \\
$^{32}$P &14.268 d & $\beta^-$ &    1.711     &0.636  & 0.332  \\
$^{34}$P &12.43 s & $\beta^-$ &    5.383     &  0.145 & 0.021  \\
$^{31}$Si &157.36 min & $\beta^-$ &    1.492     & 0.229 & 0.106   \\
Others  &&  && 1.897 &   0.022 \\
\hline
Total & &&& 4.108 & 0.733  \\
\hline
\end{tabular} 
\end{center}
\caption{Rate (\cpdt) of single scattering background events  from  \textit{in situ} produced cosmogenic isotopes, in the whole spectrum and in the region of interest for solar neutrinos [0.6--1.3] MeV.}
\label{tab:cosmo}
\end{table}

\begin{figure}
\begin{center}
\includegraphics[width=1.0\linewidth]{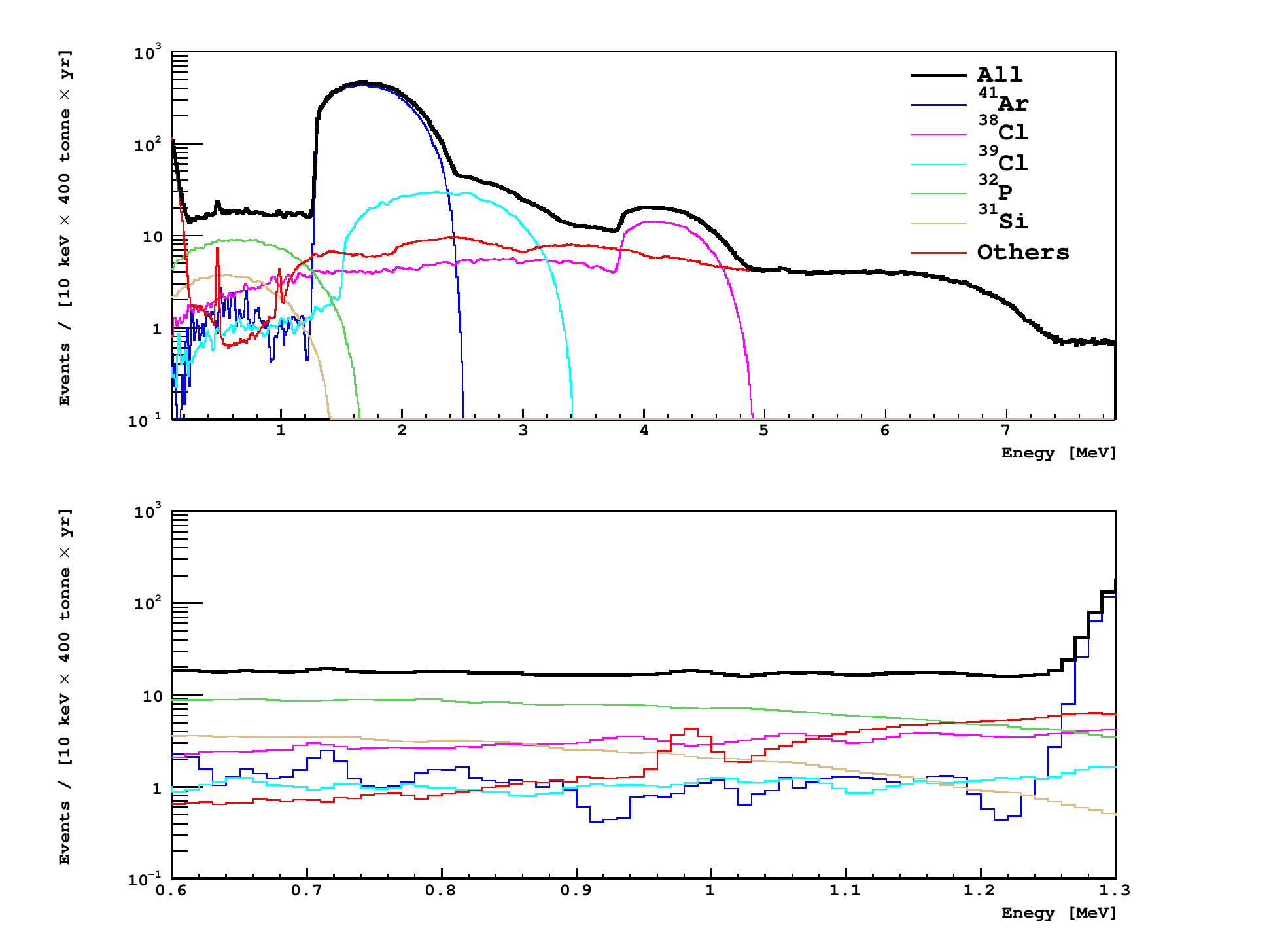}
\caption{Spectra of  cosmogenic-isotope induced backgrounds, before the multiple scattering cut, in the full energy range (top) and in the solar neutrino energy window [0.6, 1.3]~MeV (bottom).  }
\label{fig:cosmo_all_2}
\end{center}
\end{figure}

\begin{figure}
\begin{center}
\includegraphics[width=1.0\linewidth]{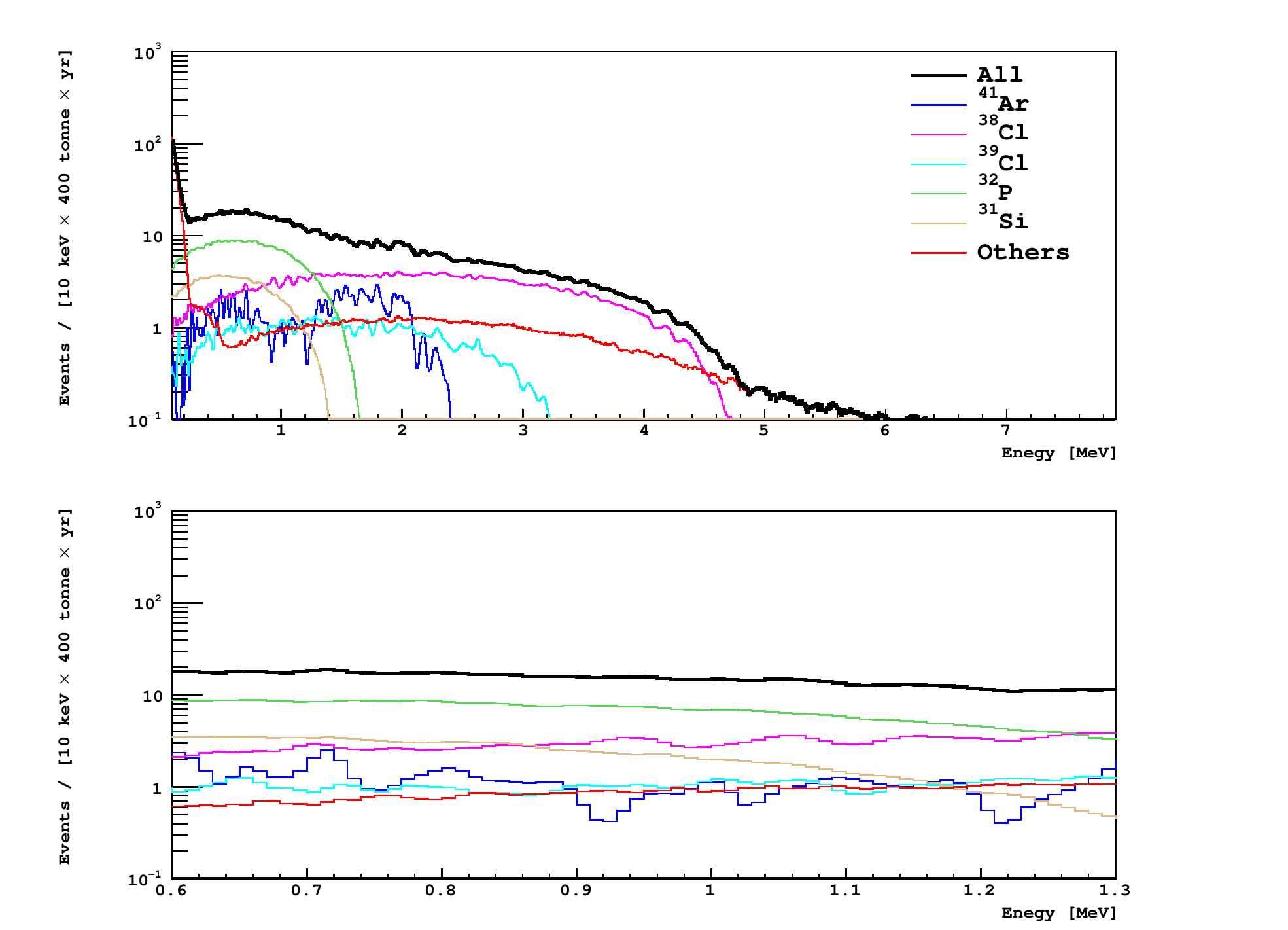}
\caption{Spectra of  cosmogenic-isotope induced backgrounds, after  the multiple scattering cut, in the full energy range (top) and in the solar neutrino energy window [0.6, 1.3]~MeV (bottom).  }
\label{fig:cosmo_all_1}
\end{center}
\end{figure}

The produced isotopes were then handed off to a GEANT4 simulation, which generated decays and tracked the decay products in the full detector geometry.  This allowed the efficiency of the multiple scattering cut to be estimated. Multiple scattering events were  conservatively  defined as events producing at least two energy deposits exceeding 10 keV, with a vertical separation exceeding 1 cm. Rejecting such events is extremely effective in eliminating $\beta^+$ decays, which have a high probability for multiple interactions due to  positron annihilation \grs, and also those $\beta^-$ decays accompanied by gamma emission. The energy region for solar neutrinos, however, is mostly populated by pure $\beta$--decay and so the multiple scattering cut rejects only $\sim$20\% of the cosmogenic background, as shown in Figures \ref{fig:cosmo_all_1} and \ref{fig:cosmo_all_2}.  

Additional rejection power would be expected by applying a delayed veto to events following a crossing muon detected in the vetoes.  However, this has not been done in the present analysis, since its efficiency depends on parameters such as the  muon rate and on the event acquisition gate length of the experiment.


The final predicted overall rate of events due to cosmogenic activities, summarized in Table \ref{tab:cosmo},  corresponds to   4.1 \cpdt\ in the entire energy range, and 0.733 \cpdt\ in the [0.6 - 1.3 MeV] energy region of interest for solar neutrinos. The neutrino signal-to-background ratio in the region of interest is thus estimated to be 6.3 (7), assuming the low (high) metallicity model. Even if we arbitrarily increase all production rates by a factor of two to allow for  uncertainties of the FLUKA production model, the signal-to-background ratio will be larger than 3.  This is ample to allow an accurate measurement of the solar neutrino components. The systematic uncertainties induced by the uncertainties on the cosmogenic isotope activities will be discussed in the next section.

As a general remark, even if the cosmogenic background does not represent a main obstacle to the solar neutrino rate measurement, as demonstrated in the next sections, a deeper experimental location, with the muon flux potentially reduced, could make this   background negligible.

%% file: Radon.tex
Contamination of the target argon mass by $^{222}$Rn may represent the most serious background problem for the solar neutrino measurements.  $^{222}$Rn emanated into the active argon from anywhere in the detector system will be distributed throughout the LAr fill  by the purification/recirculation loop. One of the goals of this work is to quantify the maximum  contamination of $^{222}$Rn that could be tolerated in a CNO neutrino experiment.

\begin{figure}
\begin{center}
\includegraphics[width=0.9\linewidth]{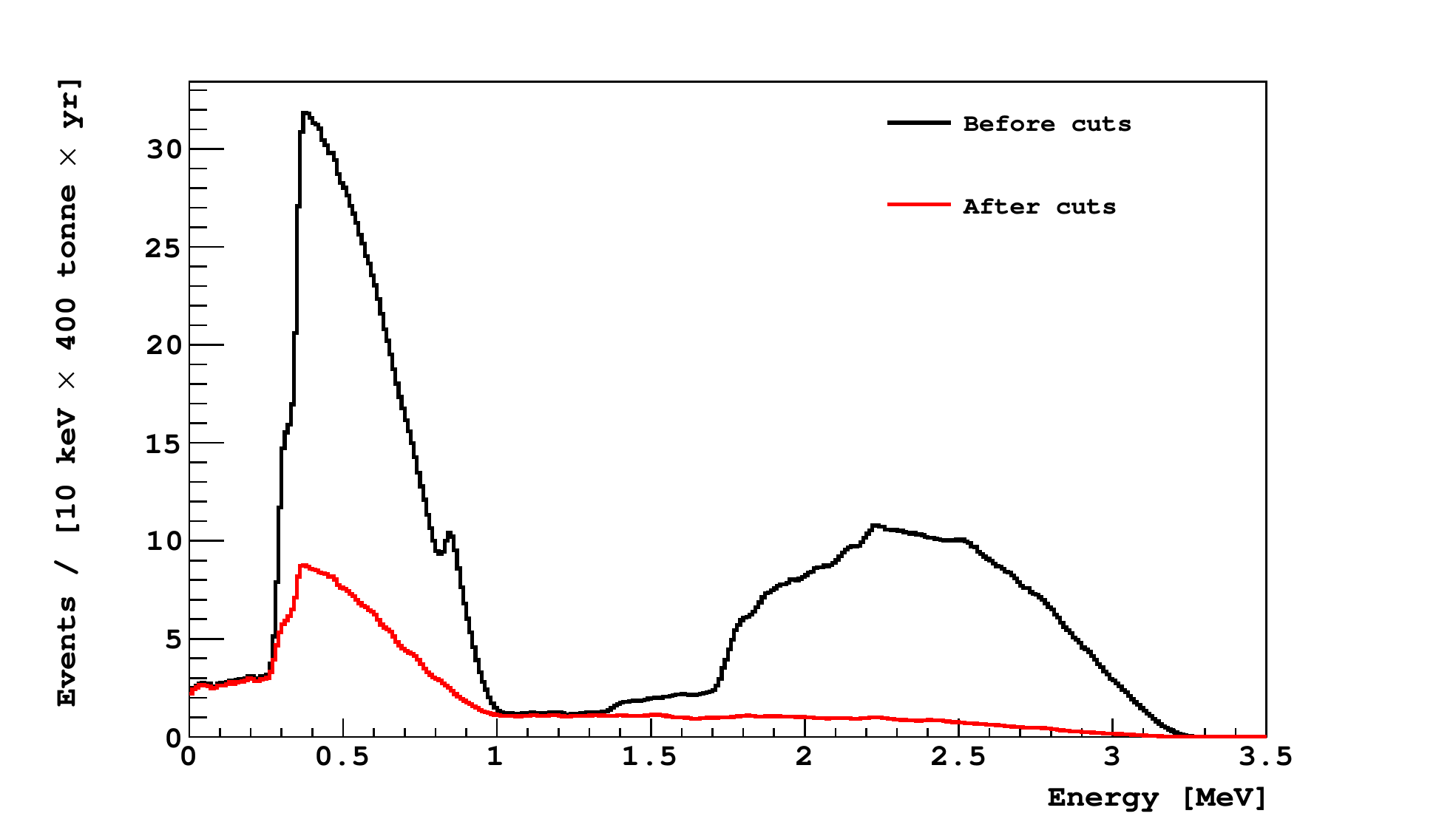}
\caption{Simulated energy spectra from  the $^{222}$Rn daughters $^{214}$Pb and $^{214}$Bi, showing the effect of the multiple interaction cut.  $^{222}$Rn activity of 10 $\mu$Bq/(100 tonne) (0.9 \cpdt) is assumed. }
\label{fig:radon}
\end{center}
\end{figure}

$^{222}$Rn $\alpha$--decay has a half life of 3.8 days, and is followed by  two $\alpha$-- ($^{218}$Po, and  $^{214}$Po) and two $\beta$--decays ($^{214}$Pb and $^{214}$Bi). The $^{222}$Rn chain segment effectively stops at the $^{210}$Pb level, due to its relatively long half life (22.3 y). In LAr,  the heavily ionizing $\alpha$ events   are  very strongly suppressed by the pulse shape discrimination: having nuclear-recoil-like pulse shape parameters \cite{Pei08},  the $\alpha$/$\beta$ discrimination factor is of the order of 10$^7$--10$^8$  \cite{Agn15}.  
Further, $\alpha$ events fall above the energy range for neutrino observation, since in LAr almost all ($\sim$85\%  \cite{Pei08}) of the $\alpha$--energy is converted into light, contrary to organic scintillators where only a small fraction ($<$10\%) is converted. 

\begin{figure}
\begin{center}
\includegraphics[width=1.0\linewidth]{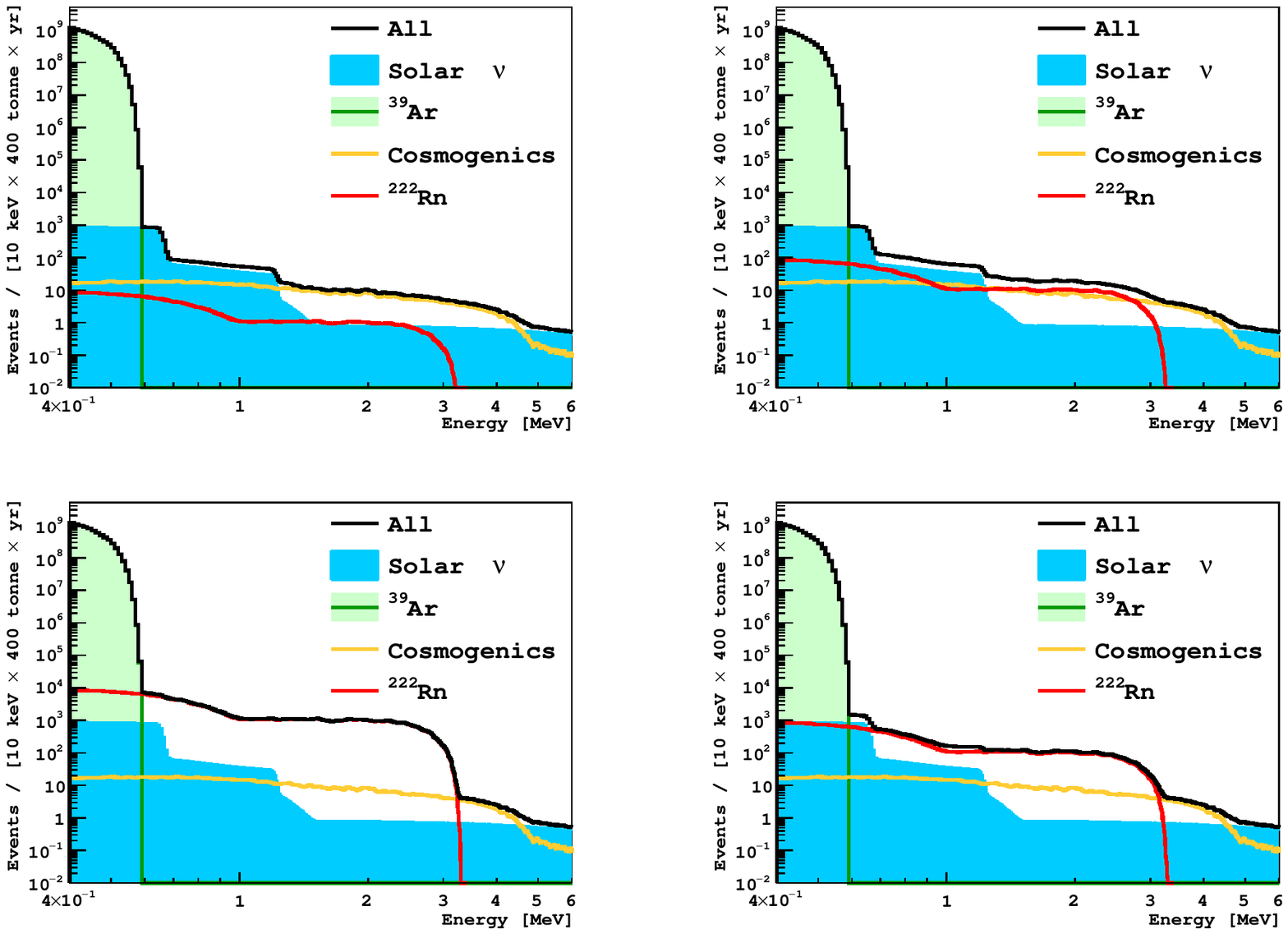}
\caption{Components of the electron recoil energy spectrum, simulated for a 4 year exposure of 100 tonnes of LAr (see legend).  Each panel shows a different assumed radon level in the LAr (clockwise from top left: 10, 100, 1,000, 10,000 $\mu$Bq/(100 tonne), corresponding to a range between 0.8 and 800 \cpdt). }
\label{fig:nuradon}
\end{center}
\end{figure}

The isotopes $^{214}$Pb and $^{214}$Bi decay by $\beta$--emission without coincident \grs\  6.3\% and 18.2\% of the time, respectively.  Those decays with coincident \grs\ can be suppressed in a two--phase \lartpc\ by eliminating events with multiple interaction sites. The effect of this cut on the total spectrum from the $^{222}$Rn decay chain, simulated with GEANT4,  is shown in Figure \ref{fig:radon}. The simulations indicate that 6.9\% (5.9\%) $^{214}$Pb ($^{214}$Bi) decays will remain with energy deposits in the  solar neutrino energy region, after the multiple--interaction cut. As a result,  ~45 $\mu$Bq/(100 tonne) (3.8 \cpdt) of $^{222}$Rn contamination would introduce a background rate equivalent to the expected solar neutrino signal.  Simulated experimental spectra for  $^{222}$Rn contaminations  ranging from  10 $\mu$Bq/(100 tonne) (0.8 \cpdt) to 10  mBq/(100 tonne) (800 \cpdt) are shown in Figure \ref{fig:nuradon}. 

The $^{222}$Rn decay rate can at least be directly measured with the \lartpc\ itself, by looking at the delayed coincidence between the $^{214}$Bi $\beta$-decay and the $^{214}$Po $\alpha$ decay ($\tau_{1/2}$ = 163 $\mu$s), a technique widely used by several low background experiments (e.g. \cite{Bel14}). Selecting $^{214}$Bi events in the [0.6, 3.3] MeV region will give a sample free of  $^{39}$Ar contamination, and containing $\sim$96\% of  the $^{214}$Bi decays. An accurate constraint on the $^{222}$Rn activity will allow the identification of the neutrino spectral shape even with high  $^{222}$Rn contaminations, as discussed in the next section.

%% file: ExternalBg.tex
The term ``external background" is used here for events from decays from radioactivity in the detector materials, such as the cryostat and the photosensors. Only the \grs\ from such activities can reach the active mass and produce an electron recoil. For the solar neutrino measurements, nuclear recoils from radio- or cosmogenic neutron scattering can be  identified and discarded using the pulse shape discrimination, with discrimination power for $\beta$/$\gamma$ vs. nuclear recoils known to be larger than 10$^7$ in  LAr \cite{Agn15}.  

To evaluate the background induced by external gamma rays affecting a solar neutrino measurement, a GEANT4 simulation was performed using $^{40}$K, $^{214}$Bi and $^{208}$Tl  
originating in the photosensors, and $^{60}$C from the cryostat.
The contaminations of  $^{40}$K, and the $^{238}$U and $^{232}$Th parents of $^{214}$Bi and $^{208}$Tl in SiPM photosensors have been reported to be smaller than 0.1 mBq/kg \cite{Ost15}. In the present work 2 mm thick silicon photosensors were simulated on the top and bottom of the TPC, corresponding to $\sim$300~kg of silicon. This gives an upper limit on the activity from the photosensors of 30 mBq. Assuming a conservative rejection factor of $10^{5}$, we expect $\sim$0.02~cpd in the [0.6--1.3]~MeV energy range, negligible with respect to the solar neutrino fluxes.

The   $^{60}$Co activity in stainless steel typical of that to be used for the cryostat has been measured by Koehler \textit{et al.} \cite{Koe04} and by  Maneschg \textit{et al.} \cite{Man08}, for different samples, with results in the range of [6.6, 45.5] mBq/kg. Assuming a 1~tonne  cryostat mass with the lowest $^{60}$Co value in the range, the overall expected $^{60}$C event rate is 570 \cpdt, with 1.7 \cpdt\ remaining  after the fiducial volume cut. The latter rate is comparable with the expected solar neutrino signal, making $^{60}$Co the most dangerous background among the external sources. Several solutions can be adopted to further reduce the $^{60}$Co contribution, such as a stronger fiducial cut, an active liquid scintillator detector surrounding the TPC working as an anti-coincidence veto (like in DarkSide-50 \cite{Agn16}),   or the identification of a stainless steel batch with lower $^{60}$Co contamination.  A titanium cryostat is a possible solution to this problem, since Ti may be produced with less radioactive contamination than stainless steel, particularly for the case of $^{60}$Co \cite{LZ15}.

Simulating the source levels just described, and after rejecting multi-sited events as discussed previously, the fraction of surviving external background events is of the order of a few tenths of a per cent as shown in Table \ref{tab:external}. In addition, the surviving events are located  close to the  TPC walls. The simulated distributions of surviving events fit exponential attenuation from the boundaries inward, with attenuation lengths varying from 3.6 to 5.1 cm depending on the source and position. As shown in Table  \ref{tab:external}, a fiducial volume cut placed 30 cm from the TPC wall would provide a rejection factor of 10$^5$, effectively suppressing the single-scatter component of external background.  

In the sensitivity study described in the next section, we assume   that the background from external sources is negligible.

\begin{table}
\begin{center}
\begin{tabular}{lcccc}
\hline
Source & Origin &Attenuation   & \multicolumn{2}{c}{Survived Fraction} \\
   &   & length  [cm]   & without FV & with FV \\
\hline
$^{40}$K & Photosensors    & 3.9 & $0.3\times10^{-2}$   &  $1\times10^{-6}$ \\
$^{214}$Bi & Photosensors & 4.2 & $1.1\times10^{-2}$   &  $9\times10^{-6}$ \\
$^{208}$Tl & Photosensors &  3.6 & $0.7\times10^{-2}$ &  $2\times10^{-6}$ \\
$^{60}$Co & Cryostat            &  5.1 & $0.1\times10^{-2}$  &  $3\times10^{-6}$ \\

\hline
\end{tabular} 
\end{center}
\caption{Fraction of events producing only a single energy deposition in the [0.6, 1.3]~MeV energy range. The last column includes a fiducial volume cut applied at 30~cm from the TPC walls.  }
\label{tab:external}
\end{table}

%% file: Sensitivity.tex
\begin{figure}
\begin{center}
\includegraphics[width=1.0\linewidth]{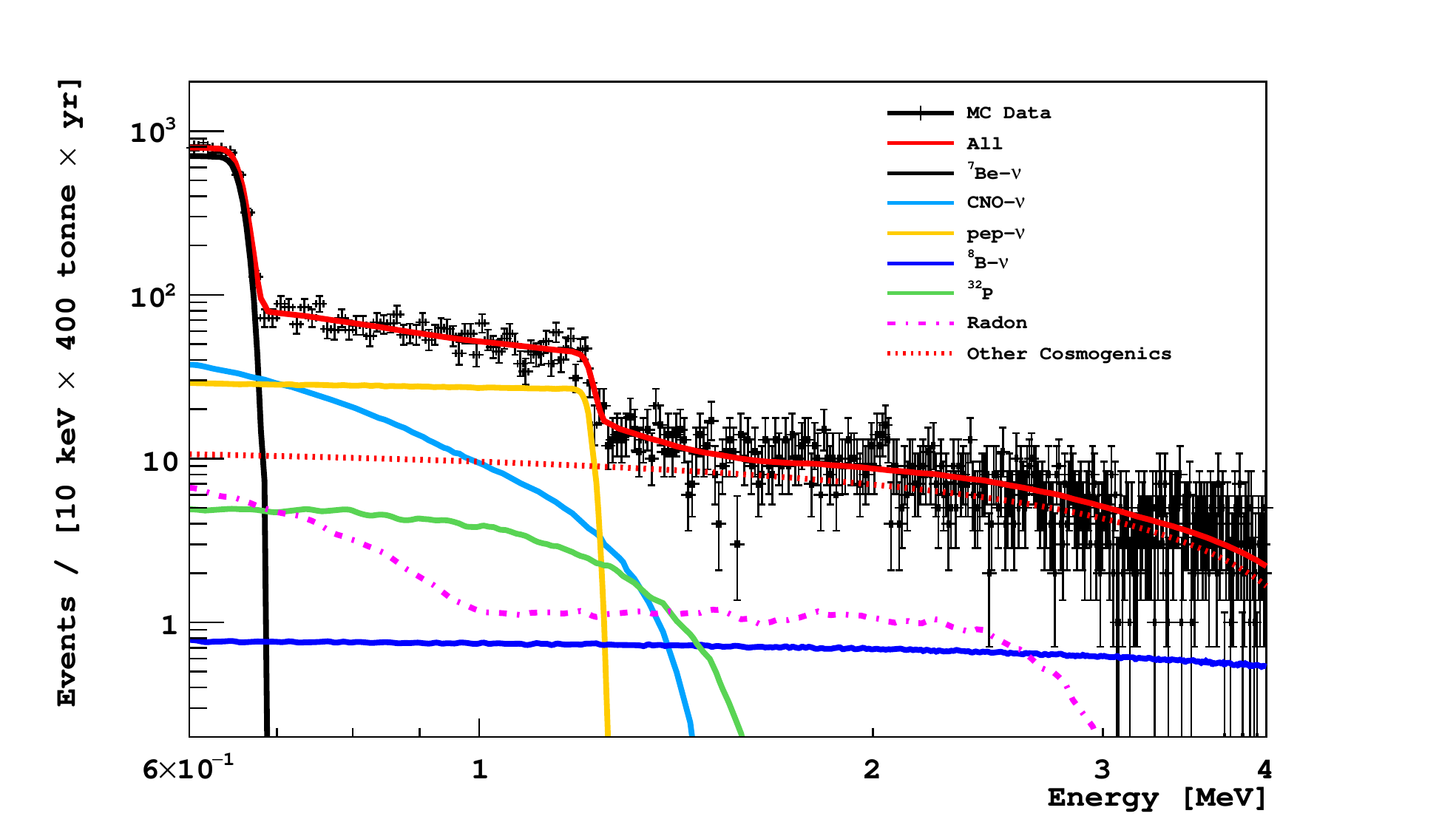}
\includegraphics[width=1.0\linewidth]{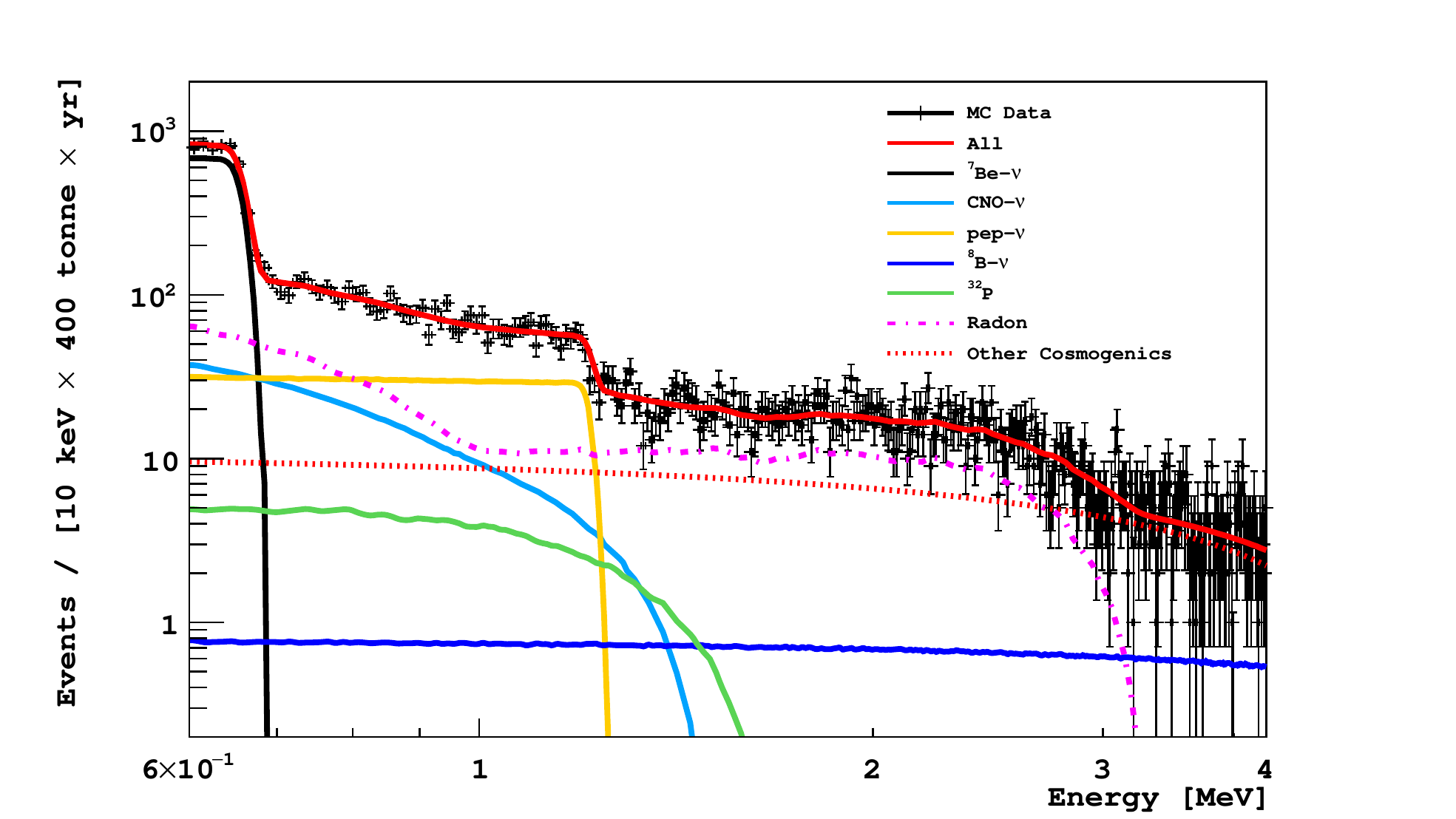}
\caption{Examples of simulated spectra and fits, assuming radon contaminations of 10 (top) and 100 (bottom) $\mu$Bq/(100 tonne) (0.8 and 8 \cpdt).  The calculations assume the low--metallicity SSM, and the cosmogenic component is modeled with a first degree polynomial, with the exception of an explicit spectrum for  $^{32}$P.  }
\label{fig:fit}
\end{center}
\end{figure}

A measurement of the solar neutrino spectral components 
relies on  identification of the  spectral shapes of the individual components. We attempt to estimate the sensitivity of a  100 tonne two--phase \lartpc\ to solar neutrinos using a toy Monte Carlo approach. For each solar metallicity model we sampled GEANT4-generated energy spectra for the various signal and background components, to produce simulated data samples. Ten thousand samples were generated with Poisson statistics corresponding to  a 400 tonne yr exposure, and assuming PE-statistics-limited detector resolution with a light yield of 6000 pe/MeV.  The external background was  neglected, since, as discussed above, it can be completely suppressed by multi-site and fiducial volume cuts. The radon contribution in the generated data samples was varied from 10 to 200~$\mu$Bq/(100 tonne) (0.8 - 16 \cpdt). 

The simulated spectra were then fit with a model designed to account for the large uncertainty of the   cosmogenic background shape. 
For the present work, only the most abundant cosmogenic contribution from $^{32}$P (see Table \ref{tab:cosmo} and Appendix \ref{appendix}) has been explicitly included in the fit function. 
The aggregate electron recoil spectrum from the remaining cosmogenics was modeled with a linear function of energy.  The parameters of the function were left free to vary in the fit which was performed with a maximum likelihood method using the RooFit package \cite{Ver03}. 

\begin{figure}
\begin{center}
\includegraphics[width=1.0\linewidth]{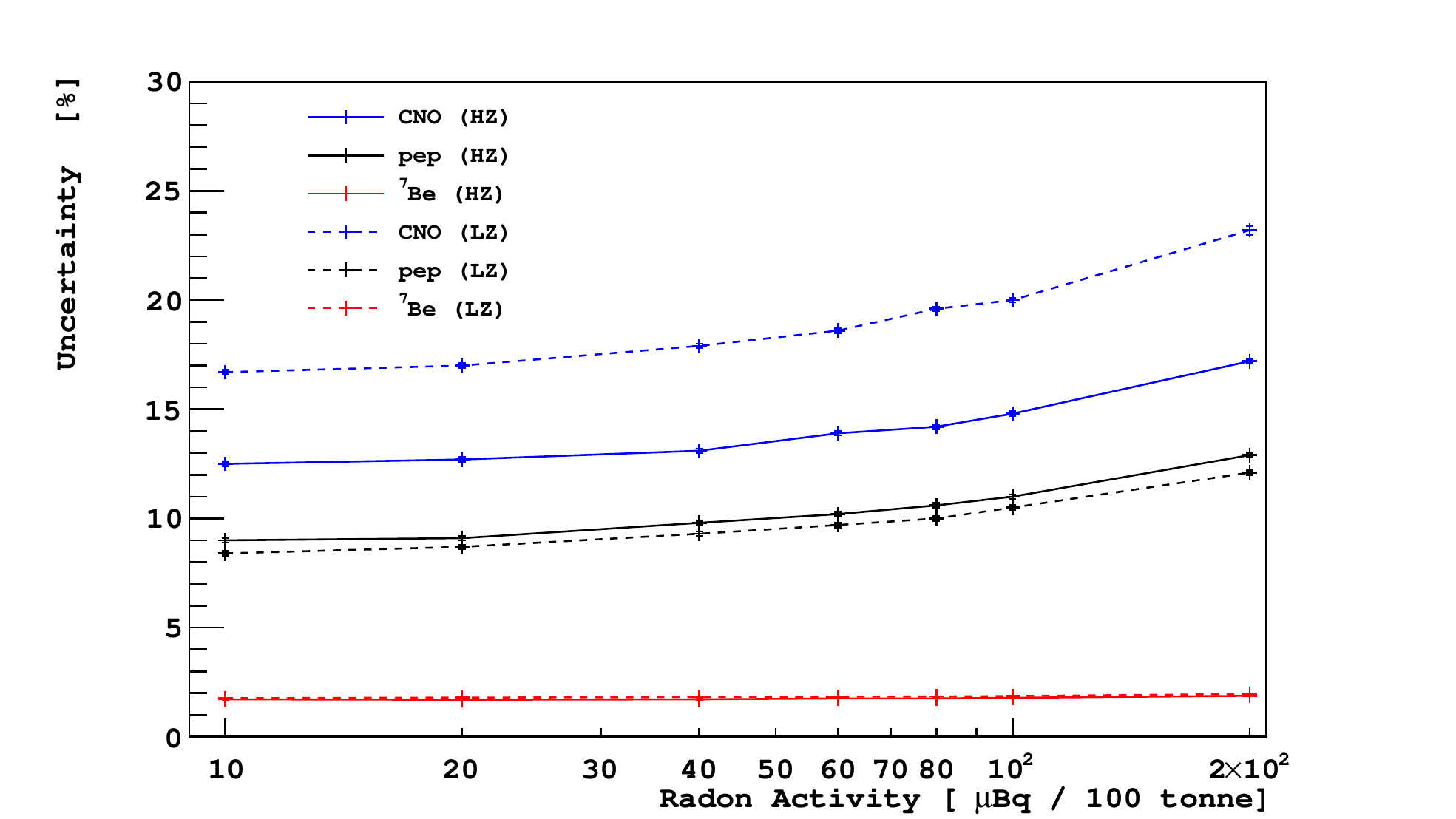}
\caption{Statistical uncertainties on the solar neutrino components, as a function of the radon activity. }
\label{fig:nu_radon}
\end{center}
\end{figure}

The radon level in the fitting model was not freely varied, but was determined from the simulated number of $^{214}$Bi-Po events in each simulated sample.  Conservatively assuming a $^{214}$Bi-Po tagging efficiency of 60\%, the intensity of the radon-daughter spectral component of the fitting model was selected with a statistical uncertainty based on the number of $^{214}$Bi-Po events seen.  The fit range extended from 0.6 to 5 MeV, so as to better constrain the background contributions. Examples  of simulated spectra and fits are shown in Figure  \ref{fig:fit} for radon contaminations between 10 and 100~$\mu$Bq/(100 tonne) (0.8 - 8 \cpdt). The resulting fitted values and uncertainties of the neutrino signal components   are given in Table \ref{tab:results}.

\begin{table}
\begin{center}
\begin{tabular}{lcccccc}
\hline
$^{222}$Rn   &  \multicolumn{3}{c}{Low Metallicity} &  \multicolumn{3}{c}{High Metallicity} \\
Activity &  $\sigma(^7Be)$ &  $\sigma(pep)$ & $\sigma(CNO)$ &  $\sigma(^7Be)$ &  $\sigma(pep)$ & $\sigma(CNO)$  \\ 
\hline

10 & 1.77 $\pm$ 0.01 & 8.4 $\pm$ 0.1 & 16.7 $\pm$ 0.1 & 1.72 $\pm$ 0.01 & 9.0 $\pm$ 0.1 & 12.5 $\pm$ 0.1 \\ 
20 & 1.80 $\pm$ 0.01 & 8.7 $\pm$ 0.1 & 17.0 $\pm$ 0.1 & 1.70 $\pm$ 0.01 & 9.1 $\pm$ 0.1 & 12.7 $\pm$ 0.1 \\ 
40 & 1.82 $\pm$ 0.01 & 9.3 $\pm$ 0.1 & 17.9 $\pm$ 0.1 & 1.72 $\pm$ 0.01 & 9.8 $\pm$ 0.1 & 13.1 $\pm$ 0.1 \\ 
60 & 1.84 $\pm$ 0.01 & 9.7 $\pm$ 0.1 & 18.6 $\pm$ 0.1 & 1.76 $\pm$ 0.01 & 10.2 $\pm$ 0.1 & 13.9 $\pm$ 0.1 \\ 
80 & 1.85 $\pm$ 0.01 & 10.0 $\pm$ 0.1 & 19.6 $\pm$ 0.1 & 1.76 $\pm$ 0.01 & 10.6 $\pm$ 0.1 & 14.2 $\pm$ 0.1 \\ 
100 & 1.87 $\pm$ 0.01 & 10.5 $\pm$ 0.1 & 20.0 $\pm$ 0.1 & 1.79 $\pm$ 0.01 & 11.0 $\pm$ 0.1 & 14.8 $\pm$ 0.1 \\ 
200 & 1.96 $\pm$ 0.01 & 12.1 $\pm$ 0.1 & 23.2 $\pm$ 0.2 & 1.88 $\pm$ 0.01 & 12.9 $\pm$ 0.1 & 17.2 $\pm$ 0.1 \\ 

\hline
\end{tabular} 
\end{center}
\caption{Solar neutrino rate uncertainties [\%] as a function of the  $^{222}$Rn  contamination [10 $\mu$Bq/(100 tonne) = 0.8 \cpdt]}
\label{tab:results}
\end{table}

The CNO spectral shape is similar to that of the low energy radon component. At radon contamination levels above 200 $\mu$Bq/(100 tonne) (16 \cpdt), the fit shows a systematic deviation from the central value of the simulated CNO component (SSM-LZ) by a few percent, implying that to guarantee a correct CNO measurement, the radon activity must be  reduced below  this level.   No such systematic deviations  are observed for   $^7$Be and \textit{pep}, whose spectral shapes have clear characteristic features. 

The  impact of a $^{85}$Kr contamination was also tested by adding  1, 10, and 100 $\mu$Bq/(100 tonne) (corresponding to 0.09, 0.9, and 9 \cpdt) of  $^{85}$Kr events  to the toy MC samples, assuming a radon contamination of 10~$\mu$Bq/(100 tonne)(0.8 \cpdt). These contamination levels resulted in an overall $^7$Be rate precisiom of $\sim$2\%, 3.5\%, and 5\%, respectively. As expected, extraction of the other neutrino components was unaffected by the $^{85}$Kr contamination. In the fitting model, the amplitude of the $^{85}$Kr 
contribution was left as a free parameter, rather than attempting to constrain it by exploiting the $^{85}$Kr-$^{85m}$Rb delayed coincidence as was done in Ref.~\cite{Agn15b}. 

The systematic uncertainty associated with the linear model of the cosmogenic background shape was also studied. For this purpose,  toy MC samples were  generated while uniformly varying each cosmogenic isotope  activity, based on the table in Appendix \ref{appendix},  by a factor between 0.5 and 2, with respect to its nominal activity. These samples were produced assuming radon contaminations of 10  and 100  $\mu$Bq/(100 tonne) (0.8 - 8 \cpdt).  The resulting variations in the extracted solar neutrino signals lie within the uncertainties reported in Table \ref{tab:results}, showing that the systematic uncertainty from the linear model assumption is unimportant. 

Other important sources of  systematic uncertainty  are expected to be the method for determining  the energy scale,  and the definition of the fiducial volume cut used to reject external background.  

The energy scale can be expected to be almost linear. In this range, Ref.~\cite{Apr87} reported a linear energy response of the LAr  ionization signal using a $^{207}$Bi source with a gridded ionization chamber. The scintillation signal is expected  to behave in the same way, being complementary to the ionization signal. In addition, the intrinsic $^{39}$Ar contamination with endpoint at 0.565 MeV,  offers a precise calibration feature to constrain the energy scale at the sub-percent level. 

Systematic uncertainty in the fiducial volume cut along the drift direction is small, since the drift velocity is known with high accuracy and precision, giving a corresponding spatial precision at the sub-millimeter  level. The dominant source of systematic uncertainty is expected to be related to the radial volume cut, since the position reconstruction in the plane orthogonal to the drift field is more complicated.
This systematic uncertainty is expected to contribute  a few percent, still not affecting the CNO and \textit{pep} measurements, which are dominated by their statistical uncertainty. It can, however, be the limiting uncertainty for  $^7$Be,  due to the high statistical precision ($\sim$1.7\%) achievable  for this component of the rate.

%% file: Discussion.tex
The present work exhibits background and signal extraction simulations indicating that a two--phase \lartpc\ with 100 tonne fiducial mass has the necessary accuracy and precision to measure  so-far unknown components of the solar neutrino spectrum.  

The two most important sources of background relevant to the proposed solar neutrino measurement are  $^{60}$Co contamination in the croystat metal,  and  $^{222}$Rn contamination within the active volume.

The $^{60}$Co issue could be solved by either applying stronger  fiducial volume cuts or by making the cryostat of titanium, which is almost free of $^{60}$Co \cite{LZ15}. Further, an external active veto as in  DarkSide-50  \cite{Agn16} would further reduce the $^{60}$Co background by vetoing  events with $^{60}$Co  gammas  detected in coincidence by the TPC and the veto.  

The required limit on the circulating $^{222}$Rn contamination, less than 200 $\mu$Bq/(100 tonne), represents the most difficult challenge. DarkSide-50 has measured a  $^{222}$Rn contamination at the level of $\sim$100 mBq/(100 tonne) \cite{Mar13}, about 3 orders of magnitude larger than what is required. However, the $^{222}$Rn emanation rate from the detector materials and the purification loop components is  expected  to provide a relatively lower contribution in a larger detector: its rate approximately scales with the ratio of the detector material surface over the LAr mass.   LAr can also be  efficiently purified   via adsorption with activated carbon columns in  both gaseous and liquid phases, as done in DarkSide-50. However, available Rn assay systems  do not have the sensitivity to monitor or verify the required level of radio-purity    \cite{Sim09}.

With 400 tonne$\times$years exposure,  precisions  at the level of $\sim$2\%,   $\sim$10\%, and  $\sim$15\%  for the  $^7$Be, \pep{} and CNO  neutrino rates,  respectively, are possible. These expectations can be compared to the best present measurements of the first two components (4.6\% and 21.6\% from Borexino), and the best upper limit for the CNO \cite{Bel14} neutrinos. 

 A CNO measurement, in particular, would provide the first direct observation of neutrinos from the CNO cycle, with the potential to discriminate between the two solar metallicity solutions. Following the same approach used in \cite{Ser13}, a measurement  of the CNO rate with 15\% precision would improve upon the current knowledge of  the C and N content in the Sun ($\pm$25\%), to the 16.5\% level.  

The simulations indicate an attainable precision for the measured $^7$Be rate which would be equivalent to an 8\% measurement of  the S17 nuclear cross section ($^7$Be(p,$\gamma$)$^8$B), which is an important input parameter in the SSM.  This is similar to the precision achieved in the direct beam--target measurement \cite{Ade11}, and would improve upon the current S17 determination from Borexino  using solar neutrino rates ($\pm$12\%) \cite{Ser13}.

In conclusion,  a large volume \lartpc\ designed for a  direct dark matter WIMP search, can also provide a rich set of physics results for solar neutrinos. Both the  precision on the $^7$Be and \pep{}   components could be significantly improved,  and potentially a first measurement of the CNO component can be achieved, if stringent $^{222}$Rn suppression levels were reached. 

%% file: Appendix.tex
\begin{table}[h]
\begin{center}
\begin{tabular}{llllc}
\hline
Isotope  &  Half life &   Decay Mode & Q-value &   Activity   \\ 
      &      &        &  [keV]   &  [cpd/(100 $\times$ tonne)] \\ 
\hline
$^{6}$He &    806.7 ms & $\beta^-$ &    3507.8     & 1.28e-01 $\pm$ 1.25e-02 \\ 
$^{8}$He &    119.1 ms & $\beta^-$ &    9670.2     & 9.68e-03 $\pm$ 3.42e-03 \\ 
$^{8}$Li &    839.9 ms & $\beta^-$ &    12975.2     & 9.55e-02 $\pm$ 1.07e-02 \\ 
$^{9}$Li &    178.3 ms & $\beta^-$ &    13606.7     & 2.06e-02 $\pm$ 4.99e-03 \\ 
$^{11}$Li &    8.75 ms & $\beta^-$ &    20231.2     & 1.21e-03 $\pm$ 1.21e-03 \\ 
$^{7}$Be &    53.22 d & EC &    861.815     & 1.02e-01 $\pm$ 1.08e-02 \\ 
$^{11}$Be &    13.76 s & $\beta^-$ &    11509.2     & 8.47e-03 $\pm$ 3.20e-03 \\ 
$^{12}$Be &    21.3 ms & $\beta^-$ &    11708     & 2.42e-03 $\pm$ 1.71e-03 \\ 
$^{8}$B &    770 ms & $\beta^+$ &    14949.8     & 9.68e-03 $\pm$ 3.42e-03 \\ 
$^{12}$B &    20.20 ms & $\beta^-$ &    13369.3     & 3.39e-02 $\pm$ 6.40e-03 \\ 
$^{13}$B &    17.36 ms & $\beta^-$ &    13437.2     & 7.26e-03 $\pm$ 2.96e-03 \\ 
$^{14}$B &    12.5 ms & $\beta^-$ &    20644     & 1.21e-03 $\pm$ 1.21e-03 \\ 
$^{9}$C &    126.5 ms & $\beta^+$ &    16494.8     & 4.84e-03 $\pm$ 2.42e-03 \\ 
$^{10}$C &    19.290 s & $\beta^+$ &    2929.62     & 8.47e-03 $\pm$ 3.20e-03 \\ 
$^{11}$C &    1221.8 s & $\beta^+$ &    1982.4     & 5.44e-02 $\pm$ 8.11e-03 \\ 
$^{14}$C &    5700 y & $\beta^-$ &    156.475     & 8.42e-06 $\pm$ 1.11e-06 \\ 
$^{15}$C &    2.449 s & $\beta^-$ &    9771.7     & 1.21e-02 $\pm$ 3.82e-03 \\ 
$^{16}$C &    0.747 s & $\beta^-$ &    7891.58     & 1.21e-03 $\pm$ 1.21e-03 \\ 
$^{12}$N &    11.000 ms & $\beta^+$ &    17338.1     & 1.21e-03 $\pm$ 1.21e-03 \\ 
$^{13}$N &    9.965 min & $\beta^+$ &    2220.49     & 3.63e-03 $\pm$ 2.09e-03 \\ 
$^{16}$N &    7.13 s & $\beta^-$ &    10419.1     & 3.87e-02 $\pm$ 6.84e-03 \\ 
$^{17}$N &    4.173 s & $\beta^-$ &    8680     & 1.21e-02 $\pm$ 3.82e-03 \\ 
$^{18}$N &    624 ms & $\beta^-$ &    11916.9     & 1.21e-03 $\pm$ 1.21e-03 \\ 
$^{14}$O &    70.606 s & $\beta^+$ &    5143.04     & 1.21e-03 $\pm$ 1.21e-03 \\ 
$^{15}$O &    122.24 s & $\beta^+$ &    2754     & 2.06e-02 $\pm$ 4.99e-03 \\ 
$^{19}$O &    26.88 s & $\beta^-$ &    4819.6     & 1.09e-02 $\pm$ 3.63e-03 \\ 
$^{20}$O &    13.51 s & $\beta^-$ &    2757.45     & 6.05e-03 $\pm$ 2.70e-03 \\

\hline
\end{tabular} 
\end{center}
\end{table}

\begin{table}[h]
\begin{center}
\begin{tabular}{llllc}
\hline

Isotope  &   Half Life & Decay Mode & Q-value &   Activity   \\ 
      &      &        &  [keV]   &  [cpd/(100 $\times$ tonne)] \\ 
\hline

$^{17}$F &    64.49 s & $\beta^+$ &    2760.8     & 3.63e-03 $\pm$ 2.09e-03 \\ 
$^{18}$F &    109.77 min & $\beta^+$ &    1655.5     & 4.11e-02 $\pm$ 7.05e-03 \\ 
$^{20}$F &    11.163 s & $\beta^-$ &    7024.53     & 3.99e-02 $\pm$ 6.95e-03 \\ 
$^{21}$F &    4.158 s & $\beta^-$ &    5684.2     & 1.57e-02 $\pm$ 4.36e-03 \\ 
$^{22}$F &    4.23 s & $\beta^-$ &    9543.42     & 2.42e-03 $\pm$ 1.71e-03 \\ 
$^{19}$Ne &    17.22 s & $\beta^+$ &    3238.4     & 2.42e-03 $\pm$ 1.71e-03 \\ 
$^{23}$Ne &    37.24 s & $\beta^-$ &    4375.81     & 8.47e-03 $\pm$ 3.20e-03 \\ 
$^{24}$Ne &    3.38 min & $\beta^-$ &    1994.39     & 4.84e-03 $\pm$ 2.42e-03 \\ 
$^{21}$Na &    22.49 s & EC &    3547.6     & 8.47e-03 $\pm$ 3.20e-03 \\ 
$^{22}$Na &    2.6027 y & $\beta^+$ &    2842.3     & 1.99e-02 $\pm$ 3.03e-03 \\ 
$^{24}$Na &    14.997 h & $\beta^-$ &    4146.78     & 8.34e-02 $\pm$ 1.00e-02 \\ 
$^{25}$Na &    59.1 s & $\beta^-$ &    3835     & 3.99e-02 $\pm$ 6.95e-03 \\ 
$^{26}$Na &    1.077 s & $\beta^-$ &    7503.27     & 3.63e-03 $\pm$ 2.09e-03 \\ 
$^{23}$Mg &    11.317 s & EC &    4056.1     & 1.21e-03 $\pm$ 1.21e-03 \\ 
$^{27}$Mg &    9.458 min & $\beta^-$ &    1766.84     & 7.01e-02 $\pm$ 9.21e-03 \\ 
$^{28}$Mg &    20.915 h & $\beta^-$ &    859.42     & 1.33e-02 $\pm$ 4.01e-03 \\ 
$^{25}$Al &    7.183 s & $\beta^+$ &    4276.7     & 4.84e-03 $\pm$ 2.42e-03 \\ 
$^{28}$Al &    2.2414 min & $\beta^-$ &    2862.77     & 2.23e-01 $\pm$ 1.64e-02 \\ 
$^{29}$Al &    6.56 min & $\beta^-$ &    2406.31     & 1.40e-01 $\pm$ 1.30e-02 \\ 
$^{30}$Al &    3.62 s & $\beta^-$ &    6325.68     & 2.78e-02 $\pm$ 5.80e-03 \\ 
$^{31}$Al &    644 ms & $\beta^-$ &    5205.97     & 2.42e-03 $\pm$ 1.71e-03 \\ 
$^{32}$Al &    33.0 ms & $\beta^-$ &    13020     & 1.21e-03 $\pm$ 1.21e-03 \\ 

\hline
\end{tabular} 
\end{center}
\end{table}

\begin{table}[h]
\begin{center}
\begin{tabular}{llllc}
\hline

Isotope  &   Half Life & Decay Mode & Q-value &   Activity   \\ 
      &      &        &  [keV]   &  [cpd/(100 $\times$ tonne)] \\ 
\hline
$^{27}$Si &    4.16 s & EC &    4812.36     & 4.84e-03 $\pm$ 2.42e-03 \\ 
$^{31}$Si &    157.36 min & $\beta^-$ &    1491.5     & 2.33e-01 $\pm$ 1.68e-02 \\ 
$^{32}$Si &    153 y & $\beta^-$ &    227.2     & 1.26e-03 $\pm$ 1.15e-04 \\ 
$^{33}$Si &    6.11 s & $\beta^-$ &    5845     & 1.69e-02 $\pm$ 4.53e-03 \\ 
$^{34}$Si &    2.77 s & $\beta^-$ &    4592     & 1.21e-03 $\pm$ 1.21e-03 \\ 
$^{29}$P &    4.142 s & EC &    4942.5     & 2.42e-03 $\pm$ 1.71e-03 \\ 
$^{30}$P &    2.498 min & EC &    4232.4     & 1.08e-01 $\pm$ 1.14e-02 \\ 
$^{32}$P &    14.268 d & $\beta^-$ &    1710.66     & 6.52e-01 $\pm$ 2.79e-02 \\ 
$^{33}$P &    25.35 d & $\beta^-$ &    248.5     & 6.29e-01 $\pm$ 2.72e-02 \\ 
$^{34}$P &    12.43 s & $\beta^-$ &    5382.96     & 2.06e-01 $\pm$ 1.58e-02 \\ 
$^{35}$P &    47.3 s & $\beta^-$ &    3988.4     & 8.10e-02 $\pm$ 9.90e-03 \\ 
$^{36}$P &    5.6 s & $\beta^-$ &    7122.1     & 4.84e-03 $\pm$ 2.42e-03 \\ 
$^{30}$S &    1.178 s & EC &    6138     & 1.21e-03 $\pm$ 1.21e-03 \\ 
$^{31}$S &    2.5534 s & EC &    5398.02     & 3.63e-03 $\pm$ 2.09e-03 \\ 
$^{35}$S &    87.37 d & $\beta^-$ &    167.33     & 9.74e-01 $\pm$ 3.28e-02 \\ 
$^{37}$S &    5.05 min & $\beta^-$ &    4865.13     & 2.87e-01 $\pm$ 1.86e-02 \\ 
$^{38}$S &    170.3 min & $\beta^-$ &    2937     & 8.95e-02 $\pm$ 1.04e-02 \\ 
$^{39}$S &    11.5 s & $\beta^-$ &    5338.78     & 1.21e-03 $\pm$ 1.21e-03 \\ 
$^{34}$Cl &    1.5266 s & EC &    5491.63     & 7.98e-02 $\pm$ 9.83e-03 \\ 
$^{38}$Cl &    37.230 min & $\beta^-$ &    4916.5     & 1.71e+00 $\pm$ 4.55e-02 \\ 
$^{39}$Cl &    55.6 min & $\beta^-$ &    3442     & 2.61e+00 $\pm$ 5.62e-02 \\ 
$^{40}$Cl &    1.35 min & $\beta^-$ &    7480     & 5.35e-01 $\pm$ 2.54e-02 \\ 
$^{35}$Ar &    1.7756 s &EC &    5966.1     & 3.63e-03 $\pm$ 2.09e-03 \\ 
$^{37}$Ar &    35.011 d &  EC &    813.87     & 1.48e+00 $\pm$ 4.16e-02 \\ 
$^{39}$Ar &    269 y & $\beta^-$ &    565     & 4.02e-02 $\pm$ 4.84e-04 \\ 
$^{41}$Ar &    109.61 min & $\beta^-$ &    2491.61     & 2.23e+01 $\pm$ 1.64e-01 \\ 
$^{38}$K &    7.636 min & EC &    5913.86     & 7.26e-03 $\pm$ 2.96e-03 \\ 

\hline
\end{tabular} 
\end{center}
\end{table}

%% file: paperLArSolarNu.bbl
\begin{thebibliography}{99}
\bibitem{Aria}DarkSide Collaboration, priv. comm. 2015.
\bibitem{Dav68}R.~Davis, D.~S.~Harmer, K.~C.~Hoffman, Phys. Rev. Lett. {\bf 20} (1968) 1205.
\bibitem{Ham98} W. Hampel \textit{et al.} (GALLEX Collaboration),  Phys. Lett.  {\bf B420} (1998) 114.
\bibitem{Alt00} M.~Altmann \textit{et al.} (GNO Collaboration), Phys. Lett. B. {\bf 490}, 16 (2000).
\bibitem{Abd09}J.N.~Abdurashitov \textit{et al.} (SAGE collaboration), Phys. Rev. C {\bf 80}, 015807 (2009).
\bibitem{Fuk96} Y. Fukuda \textit{et al.} (Kamiokande Collaboration), Phys. Rev. Lett. 77 (1996) 1683
\bibitem{Fuk98}Y. Fukuda \textit{et al.} (Super-Kamiokande Collaboration), Phys. Rev. Lett. {\bf 81}, 1562 (1998).
\bibitem{Aha05} B.~Aharmim \textit{et al.} (SNO Collaboration), Phys. Rev. C {\bf 72}, 055502 (2005).
\bibitem{Egu03} K. Eguchi \textit{et al.} (KamLAND Collaboration), Phys. Rev. Lett. {\bf 90}, 021802 (2003).
\bibitem{Arp08b} C.~Arpesella \textit{et al.} (Borexino Collaboration), Phys. Rev. Lett. {\bf 101}, 091302, (2008).
\bibitem{Arp08} C.~Arpesella \textit{et al.} (Borexino Collaboration), Phys. Lett. B {\bf 658}, 101 (2008).
\bibitem{Bel10b} G.~Bellini \textit{et al.} (Borexino Collaboration), Phys. Rev. D {\bf 82}, 033006 (2010).
\bibitem{Bel11} G.~Bellini \textit{et al.} (Borexino Collaboration),  Phys. Rev. Lett. 107 (2011) 141302.
\bibitem{Bel12} G.~Bellini \textit{et al.} (Borexino Collaboration), Phys. Rev. Lett. {\bf 108} (2012) 051302.
\bibitem{MSW} S.~P.~ Mikheyev and A.~Yu.~Smirnov, Sov. J. Nucl. Phys. {\bf 42} (1985) 913; L.~Wolfenstein, Phys. Rev. D {\bf 17} (1978) 2369.
\bibitem{Bel14} G.~Bellini \textit{et al.} (Borexino Collaboration), Phys. Rev. D {\bf 89} 112007 (2014).
\bibitem{Bel15} G.~Bellini \textit{et al.} (Borexino Collaboration), Nature {\bf 512} (2014) 7515.
\bibitem{Bel10} G.~Bellini \textit{et al.} (Borexino Collaboration), Phys. Lett. B {\bf 687} (2010) 299.
\bibitem{SBF09} A. M. Serenelli \textit{et al.} Ap. J. Lett. {\bf 705} (2009) L123.
\bibitem{SHP11} A. M. Serenelli, W. C. Haxton, C. Pe\~na-Garay, Ap. J. {\bf 743} (2011) 24.
\bibitem{AGS09} M. ~Asplund \textit{et al.}, Ann. Rev. Astron. Astroph. {\bf 47} (2009) 481.
\bibitem{GS98} N. ~Grevesse, A. ~J. Sauval, Spece Sci. Rev. {\bf 85} (1998) 161.

 
\bibitem{Sim09} H.~Simgen, G.~Zuzel, Appl. Radiat Isot.  {\bf 67} (2009)  922.
\bibitem{Agn15} P.~Agnes \textit{et al.} (DarkSide Collaboration) Phys. Lett. B {\bf 743}  (2015) 456.
\bibitem{Gal15} C.~Galbiati talk at the LNGS Scientific Committee, April 2015, https://agenda.infn.it/conferenceDisplay.py?confId=9608.
\bibitem{Agn15b} P.~Agnes \textit{et al.} (DarkSide Collaboration) Phys. Rev. D {\bf 93} (2016) 081101.

\bibitem{Dok90}T.~Doke, K.~Masuda, E.~Shibamura, Nucl. Instr. Meth. A {\bf 291-3} (1990) 617.
\bibitem{Lip10}W.~H.~Lippincott \textit{et al.}, Phys. Rev. C {\bf 81} (2010) 045803.
\bibitem{Sei02} G. M. Seidel, R. E. Lanou, W. Yao, NIM A {\bf 489} (2002) 189.
\bibitem{Ish97} N. Ishida  \textit{et al.},  NIM  A {\bf 384} (1997) 380.
\bibitem{Gra16} E.~Grace et J.~Nikkel,  arXiv:1502.04213 (2016).
\bibitem{Wri11}A.~Wright, "Optical Attenuation in Liquid Argon", DarkSide Document 41-v1, 2011,  http://darkside-docdb.fnal.gov/cgi-bin/ShowDocument?docid=41
\bibitem{Jon13} B.~J.~P.~Jones \textit{et al.}, JINST {bf 8} (2013) 09001.
\bibitem{Gor15}A.~Goretti, talk at "International conference on particle physics and astrophysics", Milan (Italy) 5-10 October 2015, http://indico.cfr.mephi.ru/event/2/.


\bibitem{Cat12}C.~Cattadori,  J. Phys.: Conf. Ser. {\bf 375}  042008 (2012). 
\bibitem{Bar97} A.~S.~Barabash, V.~N.~Kornoukhov, V.~E.~Jants NIM A {\bf 385} 530 (1997).

\bibitem{Leh10} B. E. Lehmann, S. N. Davis, and J. T. Fabryka-Martin, Water Resour. Res. {\bf 29} (2010)  2027.
\bibitem{Dav15}S.~Davini talk at TAUP 2015, September 2015, http://www.taup-conference.to.infn.it/2015/day1/parallel/dma/5\_davini.pdf.




\bibitem{Cri97}M.~Cribier \textit{et al.}, Astropart. Phys. {\bf 6} (1997) 129.
\bibitem{Hag00} T.~Hagner \textit{et al.}, Astropart. Phys. {\bf 14} (2000) 33.
\bibitem{Geant4}S. Agostinelli \textit{et al.}, Nucl. Instrum. Meth. A {\bf 506} (2003) 250; J.~Allison \textit{et al.}, IEEE T. Nucl. Sci. {\bf 53} No. 1 (2006) 270.
\bibitem{Fluka} T.T. B\"{o}hlen, F. Cerutti, M.P.W. Chin, A. Fass\`{o}, A. Ferrari, P.G. Ortega, A. Mairani, P.R. Sala, G. Smirnov and V. Vlachoudis, Nuclear Data Sheets 120, 211-214 (2014); A. Ferrari, P.R. Sala, A. Fass\`{o}, and J. Ranft, CERN-2005-10 (2005), INFN/TC\_05/11, SLAC-R-773.


\bibitem{Bel13} G.~Bellini \textit{et al.} (Borexino Collaboration),  JCAP {\bf 1308} (2013) 049.
\bibitem{Abe10}  S.~Abe \textit{et al.} (KamLAND Collaboration), Phys. Rev. C {\bf 81}  025807, 2010.
\bibitem{Amb95} M.~Ambrosio \textit{et al.} (MACRO Collaboration) Phys. Rev. D {\bf 52}  3793 (1995).
\bibitem{Aga10} N.~Agafonova \textit{et al.} (OPERA Collaboration), Eur. Phys. J. C {\bf 67} 25 (2010).
\bibitem{Ost15} I. Ostrovskiy \textit{et al.},  IEEE T. Nucl. Sci.  {\bf 62} (2015) 1825.

\bibitem{Pei08} P.~Peiffer \textit{et al.},  JINST {\bf 3} (2008) 08007.


\bibitem{Koe04}M.~Koehler \textit{et al.},  Appl. Radiat. Isot.  {\bf 61} 207 (2004). 
\bibitem{Man08} W. Maneschg \textit{et al.}, NIM A {\bf 593} 448  (2008).

\bibitem{LZ15} D. S. Akerib \textit{et al.}  (LZ  Collaboration) arXiv:1509.02910 (2015). 

\bibitem{Ver03}W.~Verkerke and D.~Kirkby  physics/0306116 (2003). 

\bibitem{Apr87}E.~Aprile, W.~Hsin-Min Ku, J.~Park, H.~Schwartz, Nucl. Instrum. Meth. A {\bf 261} Issue 3 (1987) 519.
\bibitem{Agn16} P.~Agnes \textit{et al.} (DarkSide Collaboration) JINST {\bf 11} 03 (2016) 03016.
\bibitem{Mar13}L.~Marini, Ph.D. thesis  "Dark Matter Direct Detection with DarkSide-50: Analysis of Early Data", Universit\`a degli Studi Roma Tre (2013). 

\bibitem{Ser13}A.~Serenelli, C.~Pena-Garay, W.~C.~Haxton,  Phys. Rev. D {\bf 87}  043001 (2013). 
\bibitem{Ade11}E.~G.~Adelberger \textit{et al.}, Rev. Mod. Phys. {\bf 83}  195 (2011). 


  \end{thebibliography}
